\documentclass[11pt,a4paper]{article}
\pdfoutput=1

\usepackage{jheppub}

\usepackage{amsmath}
\usepackage{bbm}
\usepackage{float}
\usepackage{graphicx}	
\usepackage{hyperref}
\usepackage{mathtools}
\usepackage{cancel}

\usepackage[normalem]{ulem}

\usepackage{multirow}
\usepackage{rotating}
\usepackage{subcaption}

\def\lsim{\raise0.3ex\hbox{$\;<$\kern-0.75em\raise-1.1ex
\hbox{$\sim\;$}}}
\def\gsim{\raise0.3ex\hbox{$\;>$\kern-0.75em\raise-1.1ex
\hbox{$\sim\;$}}}

\def\thetitle{ 
eV-scale sterile neutrino: A window open to non-unitarity? \\ 
 \vspace{- 6mm}
}
\title{\thetitle}
\hypersetup{pdftitle={\thetitle}}
\author{Hisakazu Minakata}
\affiliation{
Center for Neutrino Physics, Department of Physics, Virginia Tech, Blacksburg, Virginia 24061, USA \\
}

\emailAdd{hisakazu.minakata@gmail.com}
\date{\today}

\abstract{An excess observed in the accelerator neutrino experiments in the $\nu_{\mu} \rightarrow \nu_{e}$ channel at high confidence level (CL) has been interpreted as due to eV-scale sterile neutrino(s). But, it has been suffered from the problem of  ``appearance-disappearance tension'' at the similarly high CL because the measurements of the $\nu_{\mu} \rightarrow \nu_{\mu}$ channel do not observe the expected event number depletion corresponding to the sterile contribution in the appearance channel. We suggest non-unitarity as a simple and natural way of resolving the tension, which leads us to construct the non-unitary $(3+1)$ model. With reasonable estimation of the $\alpha$ parameters governing non-unitarity,  which we argue growing when more sterile states added, we perform an illustrative analysis to illuminate if the tension can be resolved in this model. We have found the unique solution with $\sin^2 2\theta_{14} \approx 0.3$, which is consistent with the (reactors + Ga) data and is stable against variation of the appearance signature. 
This solution bridges between the two high CL signatures, BEST and LSND-MiniBooNE, and implies a large neutrino-antineutrino asymmetry, given the much less anomaly indicated in the antineutrino sector. }

\begin{document} 

\maketitle

\section{Introduction} 
\label{sec:introduction} 

Among the varying proposals for possible candidate particles which characterize the ``beyond the Standard Model (SM)'' physics, sterile neutrino(s) is unique. It is SM gauge singlet and has no interaction with our SM world, see e.g., refs.~\cite{Dasgupta:2021ies,Boser:2019rta,Diaz:2019fwt,Kopp:2013vaa}. In a simple term, it may be characterized as having the highest ``alien degree'', or exotic character. This feature has a sharp contrast with particle dark matter~\cite{Bertone:2004pz}, which is also strongly believed to come from outside the SM. As weak (in strength) interactions between the dark and the ordinary matter are presumed, proliferating massive dark matter search experiments are ongoing, for example, as in~refs.~\cite{LZ:2022lsv,XENON:2023cxc,PandaX-4T:2021bab}. For sterile neutrino(s) the only way to look for them is to utilize the flavor mixing with the active three-generation neutrinos, rendering its experimental search highly nontrivial. 

The eV-scale sterile neutrino has a long history since the first experimental claim in 1996-2001 by the LSND collaboration as an interpretation of the $\bar{\nu}_{e}$ excess in their stopped pion source experiment~\cite{LSND:1996ubh,LSND:2001aii}. This period overlaps with the era of the milestone experimental reports~\cite{Super-Kamiokande:1998kpq,KamLAND:2002uet,SNO:2002tuh} coming out to establish the  neutrino-mass-embedded SM ($\nu$SM) with the three-generation neutrino masses and lepton flavor mixing~\cite{Maki:1962mu}. 
LSND was accompanied and followed by many other experimental searches, KARMEN~\cite{KARMEN:2002zcm}, MiniBooNE~\cite{MiniBooNE:2013uba}, MicroBooNE~\cite{MicroBooNE:2021tya}, and the short-baseline (SBL) reactor neutrino experiments including, DANSS~\cite{Danilov:2024fwi}, NEOS~\cite{NEOS:2016wee}, PROSPECT~\cite{PROSPECT:2020sxr}, STEREO~\cite{STEREO:2019ztb}, and Neutrino-4~\cite{Serebrov:2020kmd}. In fact, using the both $\bar{\nu}_{e}$ and $\nu_{e}$ appearance modes, MiniBooNE provided an evidence for their low-energy excess of 4.7 $\sigma$ confidence level (CL)~\cite{MiniBooNE:2018esg}. The same reference reports that the CL of the combined LSND and MiniBooNE excesses is as high as 6.0$\sigma$. On the other hand, some experimental searches report no evidence for the similar excess~\cite{KARMEN:2002zcm,MicroBooNE:2021tya}. 

Recently, the experimental landscape of sterile neutrino, or sterile-neutrino interpretation of the anomalies, becomes even more proliferated. A high-significance neutrino anomaly is reported from the Baksan Experiment on Sterile Transitions (BEST)~\cite{Barinov:2021asz,Barinov:2022wfh}, the $^{51}$Cr source experiment using the Ga target, which observed $\sim$20\% deficit of $\nu_e$ at $4 \sigma$ CL. It may be a definitive edition of the Ga source experiments, see ref.~\cite{Kaether:2010ag} for the reanalysis and summary of the earlier measurements. We notice that a careful analysis done in ref.~\cite{Berryman:2021yan} evaluates the BEST's significance higher than $5 \sigma$. However, it is pointed out that the BEST result has significant tension with the solar neutrino data~\cite{Berryman:2021yan,Giunti:2022btk}. 

Sometime ago, the Karlsruhe Tritium Neutrino experiment (KATRIN) started to constrain eV-scale sterile neutrino by using its high-precision electron spectrum measurement of tritium $\beta$ decay~\cite{KATRIN:2020dpx}. Quite recently, the latest KATRIN data based on 259 days of measurement is released~\cite{KATRIN:2024cdt,KATRIN:2025lph}, from which one can extract the following two important consequences: 
(1) In the three active and one sterile ($3+1$) scheme, the ``sterile-inverted ordering'' (one light, mostly sterile state and three heavy, mostly active states) is excluded without referring to cosmological observation. 
(2) The data excludes most of the region preferred by the Ga anomaly~\cite{Barinov:2021asz,Barinov:2022wfh,Kaether:2010ag} at 95\% CL, in the wide ranges of $\Delta m^2_{41}$, $1 \mbox{eV} ^2 \lsim \Delta m^2_{41} \lsim 10^3 \mbox{eV} ^2$. 

There is a progress in a completely different way of searching for eV-scale sterile. It was noticed~\cite{Nunokawa:2003ep} that it produces ``sterile-active'' resonance \`a la MSW~\cite{Wolfenstein:1977ue,Mikheyev:1985zog} in $\sim$TeV energy region, which can be searched for in the atmospheric neutrino observation in Neutrino Telescopes~\cite{IceCube:2014gqr,KM3Net:2016zxf}. For a global overview of the sterile-active resonance phenomenon, see ref.~\cite{Brettell:2024zok}. 
Recently, IceCube accumulated almost eleven years of data set which reveals a closed contour at 95\% CL in $\sin^2 2\theta_{24} - \Delta m^2_{41}$ plane, centered at $\sin^2 2\theta_{24} = 0.16$ and $\Delta m^2_{41} = 3.5$ eV$^2$~\cite{IceCubeCollaboration:2024nle,IceCubeCollaboration:2024dxk}, indicating a possibility of structure. 

Though we are not able to give a comprehensive discussion to understand the varying features of the above progresses, we revisit the problem of possible implications imposed by these new observations in the concluding section \ref{sec:conclusion}. 

In this paper, we address the particular problem called ``tension between the appearance and disappearance measurement''. See e.g., ref.~\cite{Dentler:2018sju} and the papers cited therein. We will present more informations in due course. In searching for the tension-easing solution, we may reveal a new form of existence of the sterile neutrinos as the SM gauge singlet fermions. Toward understanding the properties of possible ``sterile matter'' we believe it important to settle the issue of eV-scale sterile neutrino, its existence in nature or not, with the upcoming experiments~\cite{Ajimura:2017fld,JSNS2:2021hyk,MicroBooNE:2015bmn,Machado:2019oxb} in addition to the ongoing ones mentioned above. 

At least the two sets of experimental data claim anomalies with high CL, which may suggest us to take them as evidences for sterile neutrinos. The combined LSND-MiniBooNE excesses is at 6.0$\sigma$, and the BEST anomaly at $\gsim 5 \sigma$. However, so far, it does not appear to get a ticket for the discovered particle listings. 
What is the problem? Can theorists play a role? Apart from possible experimental issues on which the present author has no good understanding to comment, at least two problems are visible: 
\begin{enumerate} 

\item 
Problem of appearance - disappearance tension at $4.7 \sigma$ CL~\cite{Dentler:2018sju}, or higher~\cite{Hardin:2022muu}. 

\item 
Possible conflict with modern cosmology, 
see ref.~\cite{Archidiacono:2013xxa}. 

\end{enumerate}
The problem 1 implies, for short, the measurement in the $\nu_{\mu} \rightarrow \nu_{e}$ channel looks inconsistent with that in the $\nu_{\mu} \rightarrow \nu_{\mu}$ channel. In the next section~\ref{sec:UV-natural} we will give more account on this problem and propose our solution. 

In fact, we have had a quite interesting and encouraging experience while people tried to solve the problem 2, the tradition which we hope we could fellow. After strong~\cite{Hannestad:2013ana} or feeble~\cite{Dasgupta:2013zpn} self-interactions between sterile neutrinos is introduced to suppress the sterile equilibration in the universe, it spurred the various imaginative ideas. They include the possibility that the dark matter also feels this interaction~\cite{Dasgupta:2013zpn,Archidiacono:2014nda}, and that the tension between the local and CMB measurement of Hubble parameter is alleviated~\cite{Archidiacono:2015oma,Archidiacono:2014nda}. Even the possibility of having one fully thermalized sterile neutrino species is proposed~\cite{Archidiacono:2015oma}. For the background of this problem and more references see e.g., ref.~\cite{Dasgupta:2021ies}. 

\section{Non-unitarity: a natural direction}
\label{sec:UV-natural}

It is a general feature of the scattering $S$ matrix that when the inelastic channels are opened, e.g., in two-body scattering, they inevitably leads to presence of the elastic scattering. Due to unitarity, an imaginary part of the elastic scattering amplitude is generated in the presence of inelastic scattering, see e.g., ref.~\cite{Itzykson:1980rh}. Therefore, existence of the inelastic channels places a lower bound of the size of the elastic scattering. 

In this paper we take the simplest framework to treat the system of the three active plus one sterile neutrinos, the $(3+1)$ model, see section~\ref{sec:UV-(3+1)}. As unitarity is built-in in this model, opening the appearance oscillation channel $\nu_{\mu} \rightarrow \nu_{e}$ implies that we should see the disappearance channel signature, depletion of $\nu_{\mu} \rightarrow \nu_{\mu}$ at certain level, whose amount is calculable in the $(3+1)$ model. We are aware of an immediate objection to this statement, for which we have prepared Appendix~\ref{sec:partial-unitarity}.\footnote{
This intuitive reasoning was indeed the driving force which led the author to the non-unitarity approach. However, the readers who are skeptical about it (for good reasons)  are kindly invited to Appendix~\ref{sec:partial-unitarity}. 
}
Apparently, the disappearance measurements do not observe sufficient number of event depletion expected by unitarity, see e.g., ref.~\cite{Dentler:2018sju}. One may argue that the data do not respect unitarity, or, the sterile neutrino hypothesis embedded into the $(3+1)$ model does not describe our world. 

Nonetheless, the confidence level of the excess in the appearance mode is so high as $6.0 \sigma$, this is too significant to simply ignore, at least from a naive theorists' point of view. Then, one can ask the question: Is there any possible modification of the $(3+1)$ model such that it can resolve the appearance-disappearance tension? We think that the question is worth to raise because, to our view, this feature constitutes one of the important elements to prevent the $6 \sigma$ excess from having a certificate of the evidence for eV-scale sterile neutrino oscillations. 

Along this line of thought we are naturally invited to non-unitarity~\cite{Antusch:2006vwa,Escrihuela:2015wra,Fong:2016yyh,Fong:2017gke,Blennow:2016jkn}.\footnote{
Of course a complete theory must be unitary. By ``non-unitarity'' we mean the feature that a low energy effective theory becomes non-unitary when we cannot access to a new physics sector at high or low energies. For a concrete example see Appendix~\ref{sec:non-unitary-3+1}. }
The appearance-disappearance tension, or ``lack of sufficient number of elastic scattering events compared to the lower bound imposed by unitarity'', sounds the ala\'rm about possible violation of the basic principle of the $S$ matrix theory. If understood in this way, this is a fundamental problem, and there exist not so many ways to resolve it, assuming that the LSND-MiniBooNE excesses and its sterile neutrino interpretation are correct. Thus, non-unitarity is a natural and the prime candidate to serve for resolving the tension. 

In this paper, we examine the question of whether non-unitarity could resolve, or at least relax, the appearance-disappearance tension within the framework of the $(3+1)$ model.  There exist enormous number of relevant references on non-unitarity. To avoid the divergence we just quote refs.~\cite{Minakata:2021nii,Arguelles:2022tki} to enter the list and for further exploration. 

\subsection{Which framework and ingredients do we need?}
\label{sec:3+1-UV-needed}

When we observe the single (dominantly) sterile neutrino $\nu_{S}$ as a real physical object, which we assume in this paper and denote it as ``visible sterile state'', the sector of charge-neutral leptons in our world consists of the three $SU(2)_{L}$ doublets and one singlet. Assuming that this world can be described by the $(3+1)$ model, we need to implement non-unitarity into the $(3+1)$ model to ease the tension between the appearance and disappearance measurements. While the problem of sterile neutrinos~\cite{Acero:2022wqg} or non-unitarity~\cite{Arguelles:2022tki} is widely discussed in the community as possible candidates for physics beyond the $\nu$SM, we need the both, ``sterile neutrino {\em and} non-unitarity'' in our setting. 

We do not know if there exists more than one visible eV-scale sterile neutrino state. If it were the case we must extend the framework of our discussion into the non-unitary $(3+2)$ or $(3+3)$ models, the possibility we do not enter in this paper. Another point which deserves our attention is the relationship between our approach and the conventional $(3+2)$ or $(3+3)$ model simulations~\cite{Sorel:2003hf,Maltoni:2007zf,Dasgupta:2021ies,Nelson:2010hz,Giunti:2015mwa,Boser:2019rta,Diaz:2019fwt,Hardin:2022muu,Acero:2022wqg}, 
we plan to give more focused discussions in section~\ref{sec:relationship}. 

For the new physics energy scale causing non-unitarity, Blennow {\it et al.}~\cite{Blennow:2016jkn} give a separate treatment of the high- and low-scale cases in the framework of non-unitary $\nu$SM. (High- and low-scale non-unitarity in our terminologies in refs.~\cite{Fong:2016yyh,Fong:2017gke} correspond, respectively, to ``non-unitarity'' and ``sterile neutrinos'' in ref.~\cite{Blennow:2016jkn}.) In this paper we concern the low-scale case because at high-scale the prevailing $SU(2)_{L} \times U(1)$ symmetry generally leads to much severer constraints, which leaves little room for our scenario to work. For example, $| \alpha_{\mu e} | \leq 6.8 \times 10^{-4}$ in Table~1~\cite{Blennow:2016jkn}. 

In this paper, for definiteness, we focus on the mass region of the visible sterile (4th state in our model) to $\Delta m^2_{41} \equiv m^2_{4} - m^2_{1} \approx (1-10)$ eV$^2$. The formulation of low-scale non-unitary $(3+1)$ model following refs.~\cite{Fong:2016yyh,Fong:2017gke} necessitates the additional $N_{s}$ sterile states whose oscillations are ``averaged out'' in regions of the large $\Delta m^2_{41}$-driven oscillations. This requires the sterile-active mass squared differences of $N_{s}$ states, roughly speaking, greater than $\sim 100$~eV$^2$. 
In section~\ref{sec:implementing-UV} we describe how non-unitarity is implemented into the $(3+1)$ model, a sketchy description which will be backed up by Appendix~\ref{sec:non-unitary-3+1}. 

\section{Non-unitary $(3+1)$ model}
\label{sec:UV-(3+1)}  

We take the simplest framework, the non-unitary $(3+1)$ model in vacuum, to examine whether non-unitarity could resolve the problem of appearance-disappearance tension. Our analysis is for the illustrative purpose only, to show the concrete way of embodying our proposal of introducing non-unitarity, thereby showing an ``existing proof'' of the easing-tension mechanism. 

Of course, we do not attempt to resolve all possible discrepancies among the experimental data mentioned in section~\ref{sec:introduction}. Then, our non-unitary $(3+1)$ model serves as a ``working framework'' to discuss this particular tension problem. As the nature of the various discrepancies, e.g., the solar-neutrino$-$BEST tension~\cite{Berryman:2021yan,Giunti:2022btk}, 
are not understood at this stage, we try not to preclude some of the experimental data from our analysis. 

\subsection{The $(3+1)$ model in vacuum}
\label{sec:3+1-vac}

In the $(3+1)$ model in vacuum, before introducing non-unitarity, the neutrino evolution can be described by the Schr\"odinger equation in the flavor basis 
\begin{eqnarray}
i \frac{d}{dx} \nu = 
\frac{1}{2E} 
U_{(3+1)} 
\text{diag} [ 0, \Delta m^2_{21}, \Delta m^2_{31}, \Delta m^2_{41} ]
U_{(3+1)} ^{\dagger} 
\nu 
\equiv 
\frac{1}{2E} H_{(3+1)} \nu 
\label{evolution-flavor-basis}
\end{eqnarray}
where $\Delta m^2_{ji} \equiv m^2_{j} - m^2_{i}$ with the Latin mass eigenstate indices $i, j$ denote the mass squared differences between the $j$-th and $i$-th eigenstate of neutrinos ($i, j=1,2,3,4$). In eq.~\eqref{evolution-flavor-basis}, $U_{(3+1)}$ denotes the $4 \times 4$ flavor mixing matrix which relates the mass eigenstate basis to the flavor basis as $( \nu_{\text{flavor}} )_{\beta} = [ U_{(3+1)} ]_{\beta i} ( \nu_{\text{mass}})_{i}$, for which we use the Greek indices $\beta, \gamma = e, \mu, \tau, S$. It is defined as 
\begin{eqnarray} 
&& 
U_{(3+1)} 
\equiv 
U_{34} (\theta_{34}, \phi_{34} )
U_{24} (\theta_{24}, \phi_{24} )
U_{14} (\theta_{14} ) 
U_{23} (\theta_{23} ) 
U_{13} (\theta_{13}, \delta ) 
U_{12} (\theta_{12} ) 
\nonumber \\
&=&
\left[
\begin{array}{cccc}
1 & 0 & 0 & 0\\
0 & 1 & 0 & 0\\
0 & 0 & c_{34} & e^{ - i \phi_{34} } s_{34} \\
0 & 0 & - e^{ i \phi_{34} } s_{34} & c_{34}
\end{array}
\right] 
\left[
\begin{array}{cccc}
1 & 0 & 0 & 0 \\
0 & c_{24} & 0 & e^{ - i \phi_{24} } s_{24} \\
0 & 0 & 1 & 0 \\
0 & - e^{ i \phi_{24} } s_{24} & 0 & c_{24} \\
\end{array}
\right] 
\left[
\begin{array}{cccc}
c_{14} & 0 & 0 & s_{14} \\
0 & 1 & 0 & 0 \\
0 & 0 & 1 & 0 \\
- s_{14} & 0 & 0 & c_{14} \\
\end{array}
\right] 
\left[
\begin{array}{cccc}
 & & & 0 \\
 & U_{(3\times3)} & & 0 \\
 & & & 0 \\
0 & 0 & 0 & 1 \\
\end{array}
\right] 
\nonumber \\
\label{(3+1)-Umatrix}
\end{eqnarray}
where the usual abbreviated notations such as $c_{34} \equiv \cos \theta_{34}$ etc. are used. The last rotation matrix in eq.~\eqref{(3+1)-Umatrix} acts only on the active state space having the block-diagonal form of $U_{(3\times3)}$, the $\nu$SM flavor mixing matrix~\cite{Maki:1962mu}, in the first $3 \times 3$ space and unity in the 4-4 element. 
To avoid obvious conflict with the cosmological data and KATRIN, we discuss only the case that the fourth dominantly-sterile state has the heaviest mass. 

To simplify our analysis framework we make a further approximation. In the region of $L/E$ in which eV-scale sterile-active oscillation is large, the atmospheric-scale oscillation has a small effect (solar-scale one is even smaller) due to the hierarchy $\Delta m^2_{31} / \Delta m^2_{41} \simeq 10^{-3}$. Therefore, we neglect the effects of atmospheric- and solar-scale oscillations in our analysis. This can be done by setting $\Delta m^2_{31} = \Delta m^2_{21} =0$. It then implies that we can set $U_{(3\times3)} = 1$. Then, the flavor basis Hamiltonian in vacuum takes the much simplified form than the one in eq.~\eqref{evolution-flavor-basis}, 
$H_{\text{flavor}} = \frac{1}{2E} U \text{diag} [ 0, 0, 0, \Delta m^2_{41} ] U^{\dagger}$, where 
\begin{eqnarray} 
U &\equiv&
U_{34} (\theta_{34}, \phi_{34} )
U_{24} (\theta_{24}, \phi_{24} )
U_{14} (\theta_{14} ) 
\nonumber \\
&=& 
\left[
\begin{array}{cccc}
c_{14} & 0 & 0 & s_{14} \\
- e^{ - i \phi_{24} } s_{24} s_{14} & c_{24} & 0 & e^{ - i \phi_{24} } s_{24} c_{14} \\
- e^{ - i \phi_{34} } s_{34} c_{24} s_{14} & - e^{ i \phi_{24} } e^{ - i \phi_{34} } s_{34} s_{24} & 
c_{34} & e^{ - i \phi_{34} } s_{34} c_{24} c_{14} \\
- c_{24} s_{14} c_{34} & - e^{ i \phi_{24} } s_{24} c_{34} & 
- e^{ i \phi_{34} } s_{34} & c_{24} c_{14} c_{34} \\
\end{array}
\right]. 
\label{U-def}
\end{eqnarray}
The oscillation probability in vacuum can be calculated via the conventional way. 

In our analysis in this paper, we mostly concern the probabilities $P(\nu_\mu \rightarrow \nu_e)$ and $P(\nu_\mu \rightarrow \nu_\mu)$, with the data taken by the LSND and MiniBooNE experiments, and possibly others, for the former, and the accelerator long-baseline (LBL) and atmospheric neutrino measurements for the latter. For LSND and MiniBooNE the vacuum approximation should be excellent. For MINOS with the baseline $L=735$ km, for example, the matter effect exists, but a numerical examination shows that the vacuum approximation gives a reasonable first-order estimation of the probability. 
In fact, in certain perturbative frameworks such as the ones in ref.~\cite{Cervera:2000kp,Asano:2011nj}, one can give a general argument that the matter effect is absent to the first order in the expansion and it comes in only at the second order into $P(\nu_\mu \rightarrow \nu_e)$ and $1 - P(\nu_\mu \rightarrow \nu_\mu)$, the phenomenon called the ``matter hesitation''~\cite{Minakata:2017ahk}. Then, we can rely on the vacuum approximation for our purpose of performing the illustrative analysis. 

\subsection{Implementing non-unitarity into the $(3+1)$ model with $\alpha$ parametrization of the $N$ matrix}
\label{sec:implementing-UV}

We follow ref.~\cite{Fong:2016yyh} to implement non-unitarity into the $(3+1)$ model and describe the key points of the construction in Appendix~\ref{sec:non-unitary-3+1}. We start from a mother theory, the $(3+1+N_{s})$ model, with the $(4+N_{s}) \times (4+N_{s})$ unitary mixing matrix ${\bf U}$ defined as 
\begin{eqnarray} 
{\bf U} = 
\left[
\begin{array}{cc}
N & W \\
Z & V \\
\end{array}
\right]. 
\label{bfU-(3+N)}
\end{eqnarray}
$N_{s}$ denotes the number of ``oscillation-averaged-out'' sterile states, see Appendix~\ref{sec:non-unitary-3+1}. In eq.~\eqref{bfU-(3+N)} $N$ ($V$) denotes the visible sector $4 \times 4$ (averaged-out sterile sector $N_{s} \times N_{s}$) non-unitary flavor mixing matrix. $W$ and $Z$ are the transition matrices which bridge between the active and sterile subspaces, and have the appropriate rectangular shapes. 
In this formalism $N$ is simply characterized as the non-unitary $4 \times 4$ matrix, and there is no easy way to write the $N$ matrix elements in terms of the $W$'s, partly because the degrees of freedom of the $W$ matrix are much larger than $N$'s at large $N_{s}$~\cite{Fong:2016yyh,Fong:2017gke}. In these references we have expressed our preference of generic, large $N_{s}$ treatment as a natural picture of sterile matter in the non-unitarity approach. It can go beyond the explicit $N_{s}=1$ or 2 models, allowing sterile-sector model independent analyses. 

Here, we simply parametrize the non-unitary $N$ matrix by using so called the $\alpha$ parametrization~\cite{Escrihuela:2015wra}, which originates in the early references~\cite{Okubo:1961jc,Schechter:1980gr}, 
\begin{eqnarray}
N = \left( \bf{1} - \alpha \right) U
\label{N-def}
\end{eqnarray}
with the explicit form of the $\alpha$ matrix 
\begin{eqnarray}
&&
\alpha 
\equiv 
\left[
\begin{array}{cccc}
\alpha_{e e} & 0 & 0 & 0 \\
\alpha_{\mu e} & \alpha_{\mu \mu} & 0 & 0 \\
\alpha_{\tau e} & \alpha_{\tau \mu} & \alpha_{\tau \tau} & 0 \\
\alpha_{S e} & \alpha_{S \mu} & \alpha_{S \tau} & \alpha_{S S} \\
\end{array}
\right]. 
\label{alpha-def}
\end{eqnarray}
Notice that the diagonal $\alpha_{\gamma \gamma}$ elements are real, but the off-diagonal $\alpha_{\beta \gamma}$ ($\beta \neq \gamma$) elements are complex numbers. For example, $\alpha_{\mu e} = | \alpha_{\mu e} | e^{ i \phi_{\mu e}  }$. Then we define the compact notation\footnote{
In the non-unitarity implemented $\nu$SM, the observables often contain the off-diagonal complex $\alpha$ parameters which appear in combination with CP phases that are originated in the $U$ matrix~\cite{Martinez-Soler:2018lcy,Minakata:2021nii}. } 
\begin{eqnarray}
&&
\widetilde{\alpha}_{\mu e} \equiv \alpha_{\mu e} e^{ i \phi_{24} } 
= 
| \alpha_{\mu e} | e^{ i ( \phi_{\mu e} + \phi_{24} ) }. 
\label{alpha-tilde-def}
\end{eqnarray}
In harmony with our picture of non-unitarity as a probability loss in the world of $(3+1)$ neutrino flavors we assume $0 \leq \alpha_{\beta \beta} < 1$. 

For phenomenology we need the expressions of the oscillation probabilities. While a fuller treatment is sketched in Appendix~\ref{sec:non-unitary-3+1}, we take a simplified path here. That is, we replace the unitary flavor mixing matrix $U$ in eq.~\eqref{U-def} by the non-unitary $N$ matrix in eq.~\eqref{N-def}. Fortunately, the fuller treatment will not affect in any essential way our analysis to address whether non-unitarity can resolve the appearance-disappearance tension. 

\subsection{Qualitative tension easing with the non-unitary model probabilities}
\label{sec:oscillation-P}

In our analysis we use the probability formulas valid to the first order in the $\alpha_{\beta \gamma}$ parameters. This simplifies the expressions of $P(\nu_\mu \rightarrow \nu_e)$ and $P(\nu_\mu \rightarrow \nu_\mu)$ given in 
Appendix~\ref{sec:non-unitary-3+1} to: 
\begin{eqnarray}
&&
P(\nu_\mu \rightarrow \nu_e) 
\nonumber \\ 
&=& 
\biggl\{
\left\{ 1 - 2 ( \alpha_{e e} + \alpha_{\mu \mu} ) \right\} 
s^2_{24} \sin^2 2\theta_{14} 
+ 2 s_{24} \sin 2\theta_{14} \cos 2\theta_{14} 
\mbox{Re} \left( \widetilde{\alpha}_{\mu e} \right) 
\biggr\} 
\sin^2 \left( \frac{\Delta m_{41}^2 L}{4E} \right) 
\nonumber \\
&-& 
s_{24} \sin 2\theta_{14} 
\mbox{Im} \left( \widetilde{\alpha}_{\mu e} \right) 
\sin \left( \frac{\Delta m_{41}^2 L}{2E} \right). 
\label{P-mue-alpha-1st}
\end{eqnarray}
\begin{eqnarray} 
&& 
P(\nu_\mu \rightarrow \nu_\mu) 
= 
\left( 1 - 4 \alpha_{\mu \mu} \right) 
\nonumber\\
&& 
\hspace{-8mm}
- \biggl\{ 
\left( 1 - 4 \alpha_{\mu \mu} \right) 
\left( c^2_{14} \sin^2 2\theta_{24} + s^4_{24} \sin^2 2\theta_{14} \right) 
- 4 \mbox{Re} \left( \widetilde{\alpha}_{\mu e} \right) s_{24} \sin 2\theta_{14} 
( c^2_{24} - s^2_{24} \cos 2\theta_{14} ) 
\biggr\} 
\sin^2 \left( \frac{ \Delta m^2_{41} L }{ 4E } \right). 
\nonumber\\
\label{P-mumu-alpha-1st}
\end{eqnarray}
where we have ignored the probability leaking terms, as they are are of second order in the $\alpha$ parameters, see Appendix~\ref{sec:non-unitary-3+1}. The corresponding probabilities in the anti-neutrino channel can be obtained by flipping the sign of $\mbox{Im} \left( \widetilde{\alpha}_{\mu e} \right)$. 

We now observe that non-unitarity eases the tension at the qualitative level. Assuming $\mbox{Re} \left( \widetilde{\alpha}_{\mu e} \right) > 0$, the $\alpha$ parameter term makes a positive contribution to the appearance and a subtractive contribution to the disappearance channels, $1 - P(\nu_\mu \rightarrow \nu_\mu)$, letting the sterile neutrino signal larger (smaller) in $P(\nu_\mu \rightarrow \nu_e)$ ($1 - P(\nu_\mu \rightarrow \nu_\mu)$) compared to the unitary $(3+1)$ model. This feature should contribute to relax the appearance-disappearance tension. 

Nonetheless, the key question is, of course, whether the tension-easing mechanism by non-unitarity works at a semi-quantitative level. In section~\ref{sec:app-disapp-tension} we present our analysis to reveal the answer to this question. 

\subsection{Bounds required on the sterile mixing angles and the $\alpha$ parameters } 
\label{sec:s14-24-bound} 

To carry this out, we need to know the bounds on the $(3+1)$ model parameters, in particular, $s^2_{14}$ and $s^2_{24}$. In addition, and more importantly, we have to know the bounds on the $\alpha$ parameters that describe non-unitarity, which will be addressed in sections~\ref{sec:alpha-what-is} and~\ref{sec:alpha-ee-mumu-mue}. 

Extracting the bounds on $s^2_{14}$ and $s^2_{24}$ is highly non-trivial because the existing bounds on the $(3+1)$ model parameters are derived within the framework of the {\em unitary} $(3+1)$ model. We need to know how the bounds could be derived in the framework of the {\em non-unitary} $(3+1)$ model. A careful discussion on this point to derive the bound on $s^2_{14}$ will be given in Appendix~\ref{sec:sterile-mixing-bound}. Then, it will be followed by the discussion on the $s^2_{24}$ bound, for which we will rely on the MINOS and MINOS+ analyses. 

Here we give a brief summary of the constraints on $s^2_{14}$ and $s^2_{24}$, the typical values we take in our analysis, for which we leave the details to Appendix~\ref{sec:sterile-mixing-bound}. For $s^2_{14}$ we use the following three regions as the candidate best-fits, 
\begin{eqnarray}
&&
\sin^2 2\theta_{14} 
= 0.1 
~~~~~~~~
\text{(high $\Delta m^2 \gsim 7$ eV$^2$ region in reactors + solar data)},
\nonumber \\
&& 
\sin^2 2\theta_{14} 
= 0.014 
~~~~~
\text{(best fit, reactors + solar data)},
\nonumber \\
&& 
\sin^2 2\theta_{14} 
= 0.32
~~~~~~
\text{(best fit, reactors + Ga data)}. 
\label{theta-14-input}
\end{eqnarray}
For $s^2_{24}$ we use the MINOS/MINOS+ bound~\cite{MINOS:2017cae}, $s^2_{24} \leq 10^{-2}$, and take $s^2_{24} = 10^{-2}$ as a typical nonzero value. Again, we leave some details in Appendix~\ref{sec:sterile-mixing-bound}. 


\section{What is the $\alpha$ parameter?} 
\label{sec:alpha-what-is} 

It is interesting to ask what is the $\alpha$ parameter, whose answer is likely to illuminate the structure of the theory of non-unitarity. As our non-unitary $(3+1)$ model is based on the unitary $(3+1+N_{s})$ theory, the $\alpha$ parameters which describe non-unitarity should be expressed by the parameters involved in this mother theory. In fact, one can show~\cite{Escrihuela:2015wra} that by using the so called Okubo's construction of $n \times n$ matrix~\cite{Okubo:1961jc}, the $\alpha$ parameters can be written by the sterile-active mixing angles and the associated phases. This feature is emphasized by Escrihuela {\it et al.}~\cite{Escrihuela:2015wra}, and is rooted in refs.~\cite{Okubo:1961jc,Schechter:1980gr}. 

\subsection{Okubo's construction in brief}
\label{sec:okubo-brief}

Here is a brief summary of the Okubo's construction of a unitary $n \times n$ matrix as $U^{n \times n}$, where $n$ corresponds to the number of neutral fermions in the system. For simplicity, we examine the $n=6$ case, corresponding in our case  to the three active neutrinos, plus one eV visible and two heavier ``averaged-out'' sterile neutrinos. $U^{n \times n}$ has $\frac{1}{2} n (n - 1)$ rotation angles 
and one less numbers of the associated phases. $U^{6 \times 6}$ can be written as 
\begin{eqnarray}
U^{6 \times 6} &=& 
\omega_{56} \omega_{46} \omega_{36} \omega_{26} \omega_{16} 
\cdot 
\omega_{45} \omega_{35} \omega_{25} \omega_{15} 
\cdot 
\omega_{34} \omega_{24} \omega_{14} 
\cdot 
\omega_{23} \omega_{13} 
\cdot 
\omega_{12}
\label{U-nxn}
\end{eqnarray}
where $\omega_{ij}$ denotes the $n \times n$ unit matrix apart from the replacement of the $ij$ subspace by the $2 \times 2$ rotation matrix with the angle $\theta_{ij}$, (15 of them for $n=6$) and the phases $\phi_{ij}$. 

We decompose the same $U^{6 \times 6}$ in eq.~\eqref{U-nxn} into the form $U^{6 \times 6} = U^{6-4} U^{4}$, where \footnote{
If one want to start from more familiar non-unitary $\nu$SM (three active plus three sterile neutrinos) which requires a different decomposition $U^{6 \times 6} = U^{6-3} U^{3}$, visit Appendix~\ref{sec:Okubo-construction}. }
\begin{eqnarray} 
U^{6 - 4} 
&=&
\omega_{56} \omega_{46} \omega_{36} \omega_{26} \omega_{16} 
\cdot 
\omega_{45} \omega_{35} \omega_{25} \omega_{15}, 
\nonumber \\
U^{4} 
&=&
\omega_{34} \omega_{24} \omega_{14} \cdot 
\omega_{23} \omega_{13} \cdot \omega_{12}. 
\label{U64-U4}
\end{eqnarray}
Here, the upper-left $4 \times 4$ part of the $U^{4}$ matrix is nothing but the flavor mixing matrix in the (unitary) $(3+1)$ model, see eq.~\eqref{U-def}. Then, the upper-left $4 \times 4$ sub-matrix of $U^{6 - 4}$ provides us $(1 - \alpha )$ of the non-unitary $(3+1)$ model, as defined in eq.~\eqref{N-def}. The explicit expression of the $\alpha$ matrix is calculated in Appendix~\ref{sec:Okubo-construction}. To show the point, we give the expression of $\alpha$ under the small sterile-active mixing angle approximation, to second order in $s_{ij} \ll 1$, as 
\begin{eqnarray} 
&& 
\alpha 
= 
\left[
\begin{array}{cccc}
\frac{1}{2} \left( s^2_{15} + s^2_{16} \right) & 0 & 0 & 0 \\
\hat{s}_{25} \hat{s}_{15} ^* + \hat{s}_{26} \hat{s}_{16}^* & 
\frac{1}{2} \left( s^2_{25} + s^2_{26} \right) & 0 & 0 \\
\hat{s}_{35} \hat{s}_{15}^* + \hat{s}_{36} \hat{s}_{16}^* & 
\hat{s}_{35} \hat{s}_{25} ^* + \hat{s}_{36} \hat{s}_{26}^* & 
\frac{1}{2} \left( s^2_{35} + s^2_{36} \right) & 0 \\
\hat{s}_{45} \hat{s}_{15} ^* + \hat{s}_{46} \hat{s}_{16}^* & 
\hat{s}_{45} \hat{s}_{25} ^* + \hat{s}_{46} \hat{s}_{26}^* & 
\hat{s}_{45} \hat{s}_{35} ^* + \hat{s}_{46} \hat{s}_{36}^* & 
\frac{1}{2} \left( s^2_{45} + s^2_{46} \right) \\
\end{array}
\right], 
\label{alpha-4x4}
\end{eqnarray}
where we use the simplified notation $\hat{s}_{ij} \equiv s_{ij} e^{ - i \phi_{ij} }$ and $\hat{s}_{ij}^* \equiv s_{ij} e^{ i \phi_{ij} }$. 
For brevity, we denote this way of constructing the $\alpha$ matrix as the Okubo's method.  

If we we have used the same $U^{6 \times 6}$ matrix for constructing the $\alpha$ matrix in the non-unitary $\nu$SM, $\alpha^{\text{\tiny (3x3)}}$, as done in Appendix~\ref{sec:Okubo-construction} and see ref.~\cite{Blennow:2016jkn}, we would have obtained, e.g., $\alpha^{\text{\tiny (3x3)}}_{ee} = \frac{1}{2} \left( s^2_{14} + s^2_{15} + s^2_{16} \right)$ for the ``ee'' element. Therefore, the $\alpha$ matrix~\eqref{alpha-4x4} in our non-unitary $(3+1)$ model is a different object from the one in the non-unitary $\nu$SM. 
The difference between our $4 \times 4$ $\alpha$ matrix and $\alpha^{\text{\tiny (3x3)}}$ stems from the fact that the fourth sterile state is the ``visible'' state in our non-unitary $(3+1)$ model, whereas it is an ``averaged out'' state in the non-unitary $\nu$SM. 

\subsection{Growing $\alpha$ parameters as $N_{s}$ becomes larger} 
\label{sec:grow-alpha} 

We have leant that in our setting non-unitarity originates in the averaged out sterile states and its strength can be measured by the $\alpha$ parameters. Then, it is natural to suspect that the $\alpha$ parameters grow as the number of sterile states, $N_{s}$, becomes larger. We try to sharpen this argument in the rest of this section. 

If there exists a hierarchy in the sterile mixing angles, such as $s^2_{15} \ll s^2_{16} \ll s^2_{17}$ and $s^2_{25} \ll s^2_{26} \ll s^2_{27}$ etc.~treatment of the small $N_{s} (= 1,2)$ models, should suffice. But, apparently, these small $N_{s}$ models fail to resolve the tension, see section~\ref{sec:relationship}. We are interested in large $N_{s}$ cases from the spirit of low-scale non-unitarity~\cite{Fong:2016yyh}, but now we have another motivation for it. That is, we consider the large $N_{s}$ models in seeking the solution of the tension problem. In contrast to the small $N_{s}$ cases we naturally think of the randomly, or uniformly distributed $s^2_{i J}$ ($i=1,2$) at large $N_{s}$. Using the Okubo's method, see section~\ref{sec:okubo-brief}, $\alpha_{e e}$ and $\alpha_{\mu \mu}$ are given in the general $(3+1+ N_{s})$ model by 
\begin{eqnarray} 
&&
\alpha_{ee} = (s^2_{15} + s^2_{16} + \cdot \cdot \cdot + s^2_{1, N_{s} +4})/2, 
\nonumber \\
&&
\alpha_{\mu \mu} = (s^2_{25} + s^2_{26} + \cdot \cdot \cdot + s^2_{2, N_{s} +4})/2. 
\label{alpha-ee-mumu}
\end{eqnarray}
The $\alpha$ parameters increase for larger $N_{s}$ for $s^2_{i J}$ for distributed non-hierarchically.  

Of course $s^2_{ij}$ cannot be of order unity, otherwise our world is too much populated by the sterile states. There must be a universal suppression factor to prevent too frequent contact between the sterile and ordinary matter. However, it is unlikely that this suppression factor adjusts itself depending upon how many averaged-out sterile states exists in nature. This is the basis of our argument that the $\alpha$ parameters grow when $N_{s}$ increases. 

Now, we are interested in the behavior of $| \widetilde{\alpha}_{\mu e} | = | \alpha_{\mu e} |$ as $N_{s}$ increases. Notice that the value of $| \alpha_{\mu e} |$ plays the key role in the ``tension-easing'' analysis in our paper, see section~\ref{sec:app-disapp-tension}. To our present technology, the bound on $| \alpha_{\mu e} |$ can only be derived by using the Cauchy-Schwartz inequality $| \alpha_{\mu e} | \leq 2 \sqrt{ \alpha_{ee} \alpha_{\mu \mu} }$, as will be explained in section \ref{sec:Cs-bound}. From eq.~\eqref{alpha-ee-mumu} we know that the upper bound on $| \alpha_{\mu e} |$ increases as $N_{s}$ at a rate square root of the rates of $\alpha_{e e}$ and $\alpha_{\mu \mu}$. 

To summarize: Given the underlying picture of random (or uniform) sterile-active mixing angles, $\alpha_{e e}$ and $\alpha_{\mu \mu}$ tend to increase as $N_{s}$ becomes larger, unless there exists a built-in mechanism which forces either (1) each $s^2_{1J}$ (or $s^2_{2 J}$) decrease by the same rate as increasing $N_{s}$, or (2) there exists absolute upper bound on the $\alpha$ parameters. We do not know any such mechanisms of placing the upper bound on $\alpha_{e e}$ and $\alpha_{\mu \mu}$ in some concrete models, nor from general principles, such as $(4+N_{s}) \times (4+N_{s})$ space unitarity. 

\section{$\alpha$ parameter bounds} 
\label{sec:alpha-ee-mumu-mue} 

In this section, we utilize the above Okubo construction method to estimate the bounds on $\alpha_{ee}$, $\alpha_{\mu \mu}$, and $| \alpha_{\mu e} |$ for the feasibility analysis of the tension-easing mechanism by non-unitarity, to be carried out in section~\ref{sec:app-disapp-tension}. We do not claim this analysis as a complete one, but at this stage it is the only way to test if our tension-easing mechanism could work. 

We must emphasize that the $\alpha$ parameters in the non-unitary $(3+1)$ model are completely different objects from those in the non-unitarity $\nu$SM, as we noticed in section~\ref{sec:okubo-brief}. Nonetheless, we can learn many features of extraction of the $\alpha$ parameter bounds in the latter setting. For example, Blennow {\it et al.}~\cite{Blennow:2016jkn} give a comprehensive treatment. See also refs.~\cite{Parke:2015goa,Ellis:2020hus,Coloma:2021uhq}. A part of the bounds obtained in ref.~\cite{Blennow:2016jkn} is further improved by the authors of refs.~\cite{Forero:2021azc,Blennow:2025qgd}, and summarized in ref.~\cite{Arguelles:2022tki}. As ref.~\cite{Blennow:2016jkn} presents the $\alpha$ parameter bounds at 2$\sigma$ or 95\% CL we try to follow this custom. 

\subsection{Estimation of $\alpha_{ee}$ and $\alpha_{\mu \mu}$ from normalization uncertainty} 
\label{sec:Abb-normalization}

$\alpha_{\beta \beta}$ ($\beta = e, \mu$) appear in the probabilities as the overall factors, to first order, $(1 - 4 \alpha_{ee} )$ in $P(\bar{\nu}_{e} \rightarrow \bar{\nu}_{e})$ and $(1 - 4 \alpha_{\mu \mu} )$ in $P(\nu_\mu \rightarrow \nu_\mu)$. See eqs.~\eqref{P-ee} and~\eqref{P-mumu-alpha-1st} whose last term is an order of magnitude smaller. Therefore, we estimate $\alpha_{\beta \beta}$ by using the absolute normalization uncertainties. 
For an order estimate of $\alpha_{ee}$ we use the flux normalization uncertainty of the reactor neutrino flux, 5\% referring to the value quoted for the reactor antineutrino anomaly~\cite{Mention:2011rk} assuming the Huber-Mueller flux~\cite{Mueller:2011nm,Huber:2011wv}. For $\alpha_{\mu \mu}$, a typical accelerator neutrino flux uncertainty of 10\%, see e.g., ref.~\cite{T2K:2012bge}. If we identify these errors with $4 \alpha_{ee}$ and $4 \alpha_{\mu \mu}$, we obtain $\alpha_{ee} = 1.25 \times 10^{-2}$ and $\alpha_{\mu \mu} = 2.5 \times 10^{-2}$. Therefore, the order of magnitude of the diagonal $\alpha$ parameters is likely of the order of $\sim 10^{-2}$. 

\subsection{$\alpha_{ee}$ bound} 
\label{sec:Aee-bound}

Now, we try to estimate the $\alpha_{ee}$ bound in a more solid basis. We rely on the analysis of the Bugey reactor neutrino experiment which employed the three-detector fit to obtain the limit on $\sin^2 2\theta_{14}$, see Fig.~18 in ref.~\cite{Declais:1994su}. In Table 9 they quote the absolute normalization error on the neutrino flux of 2.8\% at 1$\sigma$ CL. Using the normalization uncertainty of 5.6\% at 2$\sigma$ we have estimated $\alpha_{ee}$ via this way to obtain the upper limit $\alpha_{ee} = 1.4 \times 10^{-2}$. This bound is not far from the above value obtained using the Huber-Mueller flux. 

A natural question is: Can we estimate the $\alpha_{ee}$ bound by using the Okubo's method? Goldhagen {\it et al.}~\cite{Goldhagen:2021kxe} derived the bound on $s^2_{14}$ in the $(3+1)$ model using the solar neutrino measurement, $s^2_{14} \leq 0.0168$ at 90\% CL, 1 DOF.\footnote{
They used GS98 as the default Standard Solar model, which we follow. The term GS98 refers ref.~\cite{Grevesse:1998bj}. } 
Then, the answer to the above question is {\em yes and no.} ``No'' because this is to extract a bound on $\alpha_{ee} = \frac{1}{2} \left( s^2_{15} + s^2_{16} \right)$ from the analysis for $s^2_{14}$. But, it might be ``yes'' because in the framework of ref.~\cite{Goldhagen:2021kxe}, $P(\nu_e \rightarrow \nu_e)$ is expressed as the incoherent sum over the mass eigenstates, $k=1,2,3$ (active) to 4 (sterile), which may be extended to include more sterile states. We translate the Goldhagen {\it et al.} bound to $2 \sigma$ bound, $s^2_{14} \leq 0.0275$, assuming the gaussian error. If we take the $(3+1+ N_{s})$ model with $N_{s} = 3$, we obtain the bound $( s^2_{14} + s^2_{15} + s^2_{16} + s^2_{17} ) \leq 0.0275$ at 2$\sigma$. Conservatively, it implies $\alpha_{ee} = ( s^2_{15} + s^2_{16} + s^2_{17} ) / 2 \leq 1.38 \times 10^{-2}$ in our non-unitary $(3+1)$ model.\footnote{
In Appendix~\ref{sec:sterile-mixing-bound} and hereafter, in most cases, we show the numbers in three digits. We do this to avoid accidental accumulation of the rounding errors, and therefore, it should not be understood as that the quantity has a three-digit accuracy. } 

This bound is quite consistent with the bound derived above by using the Bugey's normalization uncertainty. However, we notice that the original solar bound $( s^2_{14} + s^2_{15} + s^2_{16} + s^2_{17} ) \leq 0.0275$ excludes the region around $s^2_{14} \simeq 0.088$ preferred by the BEST result~\cite{Barinov:2021asz,Barinov:2022wfh}. It may be a part of the tension between the solar neutrino observation~\cite{Berryman:2021yan,Giunti:2022btk} and the $^{51}$Cr source experiments~\cite{Kaether:2010ag,Barinov:2021asz,Barinov:2022wfh}.\footnote{
In section~\ref{sec:existence}, however, we mention about a possible indication of different values of $\Delta m^2_{21}$ in the neutrino and antineutrino measurements. } 
Not to preclude BEST from our analysis, we do not rely on the solar neutrino bound on $\alpha_{ee}$ but stay on the bound $\alpha_{ee} = 1.4 \times 10^{-2}$ derived from the normalization uncertainty, following the discipline mentioned in section~\ref{sec:UV-(3+1)}. 

\subsection{$\alpha_{\mu \mu}$ bound} 
\label{sec:Amumu-bound} 

The generic expression of $\alpha_{\mu \mu}$ is given in eq.~\eqref{alpha-ee-mumu}. In the simplest case of $N_{s} = 1$, we need to obtain the bound on $s^2_{25}$. Even in this case we have to analyze the three frequency system with the atmospheric $\Delta m^2$, and the two sterile $\Delta m^2$ of $ \lsim 10$ eV$^2$ and $\gsim 100$ eV$^2$. In Appendix~\ref{sec:A-mumu-MINOS}, we briefly mentioned the complexity of the system, and concluded that only the MINOS/MINOS+ group~\cite{MINOS:2017cae} is able to execute the analysis. 

Now, we turn our discussion to the SK atmospheric neutrino observation~\cite{Super-Kamiokande:2014ndf}. In their sterile analysis the SK group uses the various types of samples, fully contained sub-GeV to through-going muons whose energies span from 1 GeV to $\sim$~TeV, resulting in the fully averaged-out (Fig.~10), sterile-induced oscillations which covers $1~\mbox{GeV}^2 \leq \Delta m^2 \leq 100~\mbox{GeV}^2$. As reported in Fig.~6 in ref.~\cite{Super-Kamiokande:2014ndf}, they see 3\% downward shift of normalization when the sterile is turned on. Translating the value $| U_{\mu 4} |^2 = 0.016$ used in Fig.~6 to the SK bound $| U_{\mu 4} |^2 = 0.041$ at 90\% CL, we expect $\alpha_{\mu \mu}$ of $\lsim$a few \% level value may be suggested from  this consideration. 

Fortunately the SK analysis provides us with a way of estimating $| U_{\mu 5} |^2$. In the SK's ``sterile-vacuum'' analysis they remark that the effect of sterile states (assuming two of them) comes in into $P(\nu_\mu \rightarrow \nu_\mu)$ via the form $| U_{\mu 4} |^4 + | U_{\mu 5} |^4$, incoherent contributions from the first and the second sterile states. Then, the most conservative bound on $| U_{\mu 5} |^2$ may be obtained by assuming $| U_{\mu 4} |^2 \ll | U_{\mu 5} |^2$, which implies the upper limit $| U_{\mu 5} |^2 = 0.041$ at 90\% and $| U_{\mu 5} |^2 = 0.054$ at 99\% CL. We assume the gaussian error to obtain the $2 \sigma$ bound $| U_{\mu 5} |^2 = 0.0460$. Then, the ``conservative'' bound $\alpha_{\mu \mu} = s^2_{25} / 2 = 0.023$ at $2 \sigma$ results in our non-unitary $(3+1)$ model. 
Instead, if we take the ``democratic'' ansatz $| U_{\mu 4} |^2 = | U_{\mu 5} |^2$, we obtain $| U_{\mu 4} |^4 + | U_{\mu 5} |^4 = 2 | U_{\mu 5} |^4 = ( 4.6 \times 10^{-2} )^2$, or $| U_{\mu 5} |^2 (= | U_{\mu 4} |^2) = 3.25 \times 10^{-2}$. Then, we obtain the ``democratic'' upper bound $\alpha_{\mu \mu} = s^2_{25} / 2 = 0.0163$ at $2 \sigma$. 

To convert the $| U_{\mu 4} |^2$ bound to the one on $s^2_{24}$ there is an issue of how we should treat $\theta_{14}$. However, at least the two experimental groups, MINOS~\cite{MINOS:2017cae} and SK~\cite{Super-Kamiokande:2014ndf}, examined their simulations in detail and concluded that $\theta_{14}=0$ is a good approximation to discuss the $\nu_{\mu}$ and $\bar{\nu}_{\mu}$ disappearance events. Therefore we just assume $| U_{\mu 4} |^2 = s^2_{24}$ and $| U_{\mu 5} |^2 = s^2_{25}$, the latter based on $s^2_{25} \ll 1$, in the disappearance analysis.

Thus, we have obtained  the ``conservative'' and the ``democratic'' bounds, $\alpha_{\mu \mu} \leq 0.023$ and $\alpha_{\mu \mu} \leq 0.0163$, respectively. 

\subsection{Cauchy-Schwartz bound on $| \alpha_{\mu e} |$} 
\label{sec:Cs-bound} 

The remaining $\alpha$ parameter for which we do not know the bound is $| \alpha_{\mu e} |$. In fact, as we will see in the next section~\ref{sec:app-disapp-tension}, the external $| \alpha_{\mu e} |$ bound greatly helps in our examination of the issue of the appearance-disappearance tension. Therefore, we seek the constraint on $| \alpha_{\mu e} |$ which is placed by the framework itself, in our case the non-unitary $(3+1)$ model. 

In the non-unitary $\nu$SM, the authors of ref.~\cite{Blennow:2016jkn} derived the bound $| \alpha_{\mu e} | \leq 2.8 \times 10^{-2}$ by using the KARMEN data~\cite{KARMEN:2002zcm}. However, as KARMEN is almost identical experiment with LSND, relying mostly on the stopped pion beam, determining the parameters by a younger brother experiment to fit the elder's does not look perfectly legitimate. Therefore, we seek to find an independent method to derive the $| \alpha_{\mu e} |$ bound. 

It is known that one can derive the $\alpha$ parameter bounds from a given theoretical framework by using the Cauchy-Schwartz inequality~\cite{Antusch:2006vwa}. In Table~2 in ref.~\cite{Blennow:2016jkn}, they quote the bound $\alpha_{\beta \gamma} \leq 2 \sqrt{ \alpha_{\beta \beta} \alpha_{\gamma \gamma} }$ in the non-unitary $\nu$SM, and applied it to obtain the bound $| \alpha_{\mu e} | \leq 3.2 \times 10^{-2}$ in the region $\Delta m^2 \gsim 4~\mbox{eV}^2$.\footnote{
In the case of non-unitary $\nu$SM, given the diagonal $\alpha$ parameter bounds $\alpha_{e e} \leq 2.4 \times 10^{-2}$ and $\alpha_{\mu \mu} \leq 2.2 \times 10^{-2}$ (both at 95\% CL)~\cite{Blennow:2016jkn}, $| \alpha_{\mu e} |$ bound may be obtained as $| \alpha_{\mu e} | \leq 4.6 \times 10^{-2}$. But it uses two numbers at the tip of the 95\% CL limit, and is outside of the 95\% CL region with 1 DOF. The correct bound is as above.  } 
See also ref.~\cite{Parke:2015goa}.
In our case, the non-unitary $(3+1)$ model, the Cauchy-Schwartz inequality reads 
\begin{eqnarray} 
&&
\biggl| \sum_{i=1,2,3,4} N_{\beta i} N_{\gamma i} ^* \biggr | ^2 
\leq 
\biggl( 1 - \sum_{i=1,2,3,4} \vert N_{\beta i} \vert^2 \biggr) 
\biggl( 1 - \sum_{i=1,2,3,4} \vert N_{\gamma i} \vert^2 \biggr). 
\label{Cauchy-Schwartz-def}
\end{eqnarray}
In passing we note that the left-hand side in eq.~\eqref{Cauchy-Schwartz-def} is nothing but the mis-normalization term in the probability, see eq.~\eqref{P-beta-alpha-ave-vac}. In the $\nu_\mu \rightarrow \nu_e$ channel, the left- and right-hand sides in eq.~\eqref{Cauchy-Schwartz-def} can be easily computed as, 
\begin{eqnarray} 
&&
( 1 - \alpha_{e e} )^2 | \alpha_{\mu e} |^2 
\leq 
\alpha_{ee} ( 2 - \alpha_{ee} ) 
\left( 2 \alpha_{\mu \mu} - | \alpha_{\mu e} |^2 - \alpha_{\mu \mu} ^2 \right). 
\label{Cauchy-Schwartz-bound}
\end{eqnarray}

Interestingly, the Cauchy-Schwartz bound derived in the non-unitary $\nu$SM~\cite{Blennow:2016jkn}, in its full form, has an exactly the same form as in eq.~\eqref{Cauchy-Schwartz-bound}, allowing us to use the same formula $\alpha_{\mu e} \leq 2 \sqrt{ \alpha_{ee} \alpha_{\mu \mu} }$ to obtain the bound in our non-unitary $(3+1)$ model. It appears that this property is due to our triangular parametrization of the $\alpha$ matrix in eq.~\eqref{alpha-def}. 

\subsection{Bound on $| \alpha_{\mu e} |$ through the diagonal $\alpha$ parameter bounds} 
\label{sec:diagonal-alpha}

With the bounds on $\alpha_{ee}$ and $\alpha_{\mu \mu}$ at hand, we are ready to derive the $| \alpha_{\mu e} |$ bound. For $\alpha_{ee}$ we use the value estimated by using the normalization uncertainty $\alpha_{ee} \leq 1.4 \times 10^{-2}$. For $\alpha_{\mu \mu}$, we have utilized in section~\ref{sec:Amumu-bound} the SK atmospheric neutrino analysis to obtain the bound $\alpha_{\mu \mu} \leq 0.023$ (conservative case), and $\alpha_{\mu \mu} \leq 0.0163$ (democratic case), each at 2$\sigma$ CL. Then, we obtain the Cauchy-Schwartz bound $| \alpha_{\mu e} | \leq 2 \sqrt{ \alpha_{ee} \alpha_{\mu \mu} }$ at 2$\sigma$ CL: 
\begin{eqnarray} 
&&
| \alpha_{\mu e} | \leq 2.54 \times 10^{-2} 
~~~~~~\text{(Conservative)}, 
\nonumber \\
&&
| \alpha_{\mu e} | \leq 2.13 \times 10^{-2} 
~~~~~~\text{(Democratic)}. 
\label{Amue-bound}
\end{eqnarray}
We use these bounds in our analysis in section~\ref{sec:app-disapp-tension}. One may ask which bound, conservative or democratic in the above, is our ``official'' one? We consider none of them official. As a matter of fact, we will find a posteriori in section~\ref{sec:app-disapp-tension}, that our tension easing solutions obtained there have the value $s^2_{24} \approx 10^{-2}$, such that it is located mid between the conservative and democratic bounds above. 

\section{Can non-unitarity relax the appearance-disappearance tension?}
\label{sec:app-disapp-tension} 

Now we address the question of whether introduction of non-unitarity can relax the appearance-disappearance tension in a sufficient way to make the model phenomenologically viable. To answer this question we seek to find the consistent solution of the appearance and disappearance equations (see eq.~\eqref{leading-model} below for the simplest case) under the constraints on the sterile-active mixing angles and the relevant $\alpha$ parameters in the non-unitary $(3+1)$ model. The former is estimated in sections~\ref{sec:s14-bound} and~\ref{sec:s24-bound}, and we use our $\alpha$ parameter bounds summarized in section~\ref{sec:diagonal-alpha}, with the most important ones given in eq.~\eqref{Amue-bound}. 
Nonetheless, our analysis is at the level of illustrative purpose, i.e., to present an existence proof of the successful tension easing mechanism.

\subsection{The leading-order model} 
\label{sec:leading-order-model}

In this paper our analysis will be carried out under the various simplifying assumptions: 
\begin{itemize}

\item 
The expressions of $P(\nu_\mu \rightarrow \nu_e)$ and $P(\nu_\mu \rightarrow \nu_\mu)$ in eqs.~\eqref{P-mue-alpha-1st} and \eqref{P-mumu-alpha-1st} contains $\alpha_{e e}$ and $\alpha_{\mu \mu}$ as well as $\widetilde{\alpha}_{\mu e} = | \alpha_{\mu e} | e^{ i ( \phi_{\mu e} + \phi_{24} ) }$. Lacking any hints from the experiments we assume that all the phase parameters vanish, $\phi_{\mu e} = \phi_{24} = 0$, or $\mbox{Im} \left( \widetilde{\alpha}_{\mu e} \right) = 0$.  

\item 
We assume that $\alpha_{e e}=\alpha_{\mu \mu}=0$. This is a reasonable start setting, given the upper bounds of the order of $10^{-2}$ for the both parameters. 

\end{itemize}
Then, our analysis will proceed via the following two-step strategy: 
(1) By setting the order unity coefficients, such as $\cos 2\theta_{24}$ and $\cos 2\theta_{14}$, equal to unity in eqs.~\eqref{P-mue-alpha-1st} and \eqref{P-mumu-alpha-1st}, we define the ``leading-order model'' which, we hope, successfully captures the key features of the system. 
(2) After solving the leading-order model we show that the obtained solution is stable against inclusion of the first order corrections. 

Following the above construction the leading-order model reads: 
\begin{eqnarray}
P(\nu_\mu \rightarrow \nu_e) 
&=&
\left[ 
s^2_{24} \sin^2 2\theta_{14} + 2 s_{24} \sin 2\theta_{14} 
\mbox{Re} \left( \widetilde{\alpha}_{\mu e} \right)
\right] 
\sin^2 \left( \frac{\Delta m_{41}^2 L}{4E} \right),  
\nonumber \\
1 - P(\nu_\mu \rightarrow \nu_\mu) 
&=& 
4 \left[ 
s^2_{24} - s_{24} \sin 2\theta_{14} \mbox{Re} \left( \widetilde{\alpha}_{\mu e} \right) 
\right] 
\sin^2 \left( \frac{ \Delta m^2_{41} x }{ 4E } \right).  
\label{leading-model}
\end{eqnarray}
In the second line of $P(\nu_\mu \rightarrow \nu_\mu)$ in eq.~\eqref{P-mumu-alpha-1st} we have ignored the $s^4_{24} \sin^2 2\theta_{14}$ term because it is tiny, $\lsim 10^{-4}$. In this setting, the easing mechanism for the appearance-disappearance tension relies on the unique parameter, $\mbox{Re} \left( \widetilde{\alpha}_{\mu e} \right) = | \alpha_{\mu e} |$, as we have ignored the CP phases. 

\subsection{Parameters used in the analysis}
\label{sec:param-used} 

In our discussions in Appendix~\ref{sec:sterile-mixing-bound} on the mixing angle bound we have focused on the particular types of the experiments to illuminate its validity in our framework of the non-unitary $(3+1)$ model. In this section we mention about how inclusion of the other relevant measurements improves the $| U_{e 4} |$ and $| U_{\mu 4} |$ determination to decide the experimental input for our analysis. 
The LSND experiment~\cite{LSND:2001aii} measures the coefficient of the $\sin^2 \left( \Delta m_{41}^2 L / 4E \right)$ in $P(\nu_\mu \rightarrow \nu_e)$, $\sin^2 2\theta_{\mu e} \equiv 4 | U_{e 4} U_{\mu 4} |^2 = s^2_{24} \sin^2 2\theta_{14}$. Including the MiniBooNE~\cite{MiniBooNE:2013uba}, KARMEN~\cite{KARMEN:2002zcm}, and the other relevant experiments, the authors of ref.~\cite{Dentler:2018sju} obtained the allowed region of $\sin^2 2\theta_{\mu e}$ in the range $2 \times 10^{-3} \lsim \sin^2 2\theta_{\mu e} \lsim 2 \times 10^{-2}$ at 99\% CL for 2 DOF. See Fig.~4 in ref.~\cite{Dentler:2018sju}. For concreteness we adopt the value $\sin^2 2\theta_{\mu e} = 6 \times 10^{-3}$ (close to the best fit) as the reference value in our analysis, and check the stability of our conclusion by allowing variation within the above range. This implies in our leading-order version of the non-unitary $(3+1)$ model 
\begin{eqnarray} 
&&
s^2_{24} \sin^2 2\theta_{14} 
+ 2 s_{24} \sin 2\theta_{14} | \alpha_{\mu e} | 
= 
6 \times 10^{-3}.  
\label{LSND-result}
\end{eqnarray}
To repeat our logic again, the right-hand side of eq.~\eqref{LSND-result} is the experimentally measured coefficient of the $\sin^2 \left( \Delta m_{41}^2 L / 4E \right)$, and the left-hand side the theoretical expression of the same quantity in our non-unitary $(3+1)$ model. 

The global analysis of the disappearance measurement of $P(\nu_\mu \rightarrow \nu_\mu)$ and $P(\bar{\nu}_\mu \rightarrow \bar{\nu}_\mu)$ to constrain $| U_{\mu 4} |^2 = s^2_{24} c^2_{14}$ is also carried out in ref.~\cite{Dentler:2018sju} by including the data not only from MINOS/MINOS+ but also SK, IceCube, IceCube-Deep-Core etc. It may be fair to summarize the bound they obtained (as presented in Fig.~5) as $| U_{\mu 4} |^2 \lsim 10^{-2}$ in the region $1~\mbox{eV}^2 \lsim \Delta m^2_{41} \lsim 10~\mbox{eV}^2$ at the same CL for $\sin^2 2\theta_{\mu e}$. 
As explained in section~\ref{sec:Amumu-bound}, we set $| U_{\mu 4} |^2 = s^2_{24}$ in the disappearance analysis. 
It implies in the leading-order model, following the same logic as for eq.~\eqref{LSND-result}, 
\begin{eqnarray} 
&&
4 \left[ s^2_{24} - s_{24} \sin 2\theta_{14} | \alpha_{\mu e} | \right] 
\leq 
4 \times 10^{-2}. 
\label{MINOS-bound}
\end{eqnarray}
Since the way of how the right-hand side of eq.~\eqref{MINOS-bound} is estimated lacks a proper statistical ground, we cannot offer, for example, the $2 \sigma$ allowed region of the above value. 

\subsection{Analysis of the leading order model: Case of small $\theta_{14}$} 
\label{sec:leading-order-model-1} 

To illuminate the structure of the leading order model, we cast the model into a simple pictorial form. For convenience of our discussion we define the variables 
\begin{eqnarray} 
&&
X \equiv s_{24} \sin 2\theta_{14}, 
\hspace{10mm}
Y \equiv s_{24}, 
\hspace{10mm}
Z \equiv | \alpha_{\mu e} | > 0, 
\label{variables-def}
\end{eqnarray}
to rewrite eqs.~\eqref{LSND-result} and~\eqref{MINOS-bound} as 
\begin{eqnarray} 
&&
X^2 + 2 X Z = A, 
\nonumber \\
&& 
4 Y^2 - 4 X Z = B, 
\label{coupled-eqs}
\end{eqnarray}
where $A = 6 \times 10^{-3}$ and $B = 4 \times 10^{-2}$. In what follows we sometimes refer $A$ and $B$ as the ``appearance constant'' and ``disappearance constant'', respectively. 
We note that we have replaced the inequality in eq.~\eqref{MINOS-bound} by the equality because if $B$ becomes smaller it becomes harder to ease the tension. Therefore, eq.~\eqref{coupled-eqs} is the easiest case for us to be able to relax the tension. 

By eliminating $X Z$ from eq.~\eqref{coupled-eqs} we obtain the $Z$ independent ellipse equation 
\begin{eqnarray} 
&&
\frac{ X^2 }{ \left( \sqrt{ A + \frac{ B }{2} } \right) ^2 } 
+ \frac{ Y^2 } { \left\{ \sqrt{ \frac{1}{2} \left( A + \frac{ B }{2} \right) } \right\}^2 } 
= 1 
\label{ellipse-eq}
\end{eqnarray}
with the lengths of the major and minor axes $\sqrt{ A + \frac{ B }{2} } = 0.161$ and $\sqrt{ \frac{1}{2} \left( A + \frac{ B }{2} \right) } = 0.114$, respectively. This ellipse is independent of $Z$, and hence of the $\alpha$ parameter. If the crossing point $(X_{c}, Y_{c})$ with the straight line $Y = (\sin 2\theta_{14})^{-1} X$ exists at the right place, we have the favorable ``easing tension'' solution. 

Now we examine small $\sin 2\theta_{14}$ case, $(\sin 2\theta_{14})^{-1} \gg 1$. An example of such case is provided by the best fit point of the reactor-solar data implies $\sin^2 2\theta_{14} = 0.014$ which means $s_{14} = 0.0593$ and $(\sin 2\theta_{14})^{-1} = 8.45 \gg 1$. Because the slope of the straight line is large, the crossing point is close to the $Y$ axis. Therefore, $Y_{c} \simeq \sqrt{ \frac{1}{2} \left( A + \frac{ B }{2} \right) } = 0.114$, which is a quite reasonable value for $s_{24}$. Then, $X_{c} = Y_{c} \sin 2\theta_{14}$ is an order of magnitude smaller than $Y_{c}$. 
Then, in a good approximation the second line in eq.~\eqref{coupled-eqs} gives 
\begin{eqnarray} 
&&
XZ \simeq Y_{c}^2 - \frac{B}{4} 
= 
\frac{1}{2} A, 
\label{XZ-sol}
\end{eqnarray}
which means $Z = \frac{1}{2} \frac{A}{X}$. Using $X=X_{c}$ we obtain 
\begin{eqnarray} 
&&
Z = | \alpha_{\mu e} | 
= \frac{1}{2} \frac{A}{ Y_{c} \sin 2\theta_{14} } 
= 
\frac{ 2.63 \times 10^{-2} }{ \sin 2\theta_{14} } 
\leq 
2.54 \times 10^{-2}. 
\label{CS-small-14}
\end{eqnarray}
In the last inequality we have used the bound on $| \alpha_{\mu e} |$ (conservative case) obtained in section~\ref{sec:diagonal-alpha}. Equation~\eqref{CS-small-14} means that $\sin 2\theta_{14} \sim 1$, which does not qualify as a small $\theta_{14}$ solution. In fact, $\sin 2\theta_{14}$ exceed unity for this particular value of $A$. If we use the tighter constraint $| \alpha_{\mu e} | \leq 2.13 \times 10^{-2}$ (democratic) the situation becomes worse, as $\sin 2\theta_{14}$ becomes larger. Thus, we can conclude quite generally from the pictorially-drawn leading-order model that no easing tension solution can be found for a small $\theta_{14}$, $(\sin 2\theta_{14})^{-1} \gg 1$. 

\subsection{Analysis of the leading-order model: Case of large $\theta_{14}$} 
\label{sec:leading-order-model-2} 

In the case of large $\theta_{14}$, e.g., $\sin 2\theta_{14} = 0.32$ which is the best fit to the reactor + Ga data mentioned in section~\ref{sec:s14-bound}, we can no longer use the ``steep slope'' approximation. Therefore, we use the alternative method to solve the leading-order model. 

We first discuss the case of saturated Cauchy-Schwartz bound, $Z = | \alpha_{\mu e} | = 2.54 \times 10^{-2}$ (conservative). For a given $Z$ we can solve the first line of eq.~\eqref{coupled-eqs} with the solution 
\begin{eqnarray} 
&&
X_{0} = 
\left[ - Z + \sqrt{ Z^2 + A } \right], 
\label{X0-sol}
\end{eqnarray}
where we have picked the plus sign because $X>0$. Then the solution to the second equation is given by 
\begin{eqnarray} 
&&
Y_{0}^2 = Z X_{0} + \frac{B}{4} 
= 
Z \left[ - Z + \sqrt{ Z^2 + A } \right] + \frac{B}{4}. 
\label{Y0-sol}
\end{eqnarray}
For the given values $A = 6 \times 10^{-3}$ and $B = 4 \times 10^{-2}$, we obtain $X_{0} = 5.61 \times 10^{-2}$ and $Y_{0} = 0.107$, which means $X_{0} / Y_{0} = \sin 2\theta_{14} = 0.524$. Or, $\sin^2 2\theta_{14} = 0.275$, the value reasonably close to $\sin^2 2\theta_{14} = 0.32$, the best fit to the reactors + Ga data mentioned in section~\ref{sec:s14-bound}. In fact, the value $\sin^2 2\theta_{14} = 0.275$ is within the allowed islands in the combined analysis of the reactors and Ga data at $2 \sigma$ CL~\cite{Berryman:2021yan}. In passing we remark that the value of $\theta_{24}$, $s_{24} = Y_{0} = 0.107$, is quite reasonable. 

Now we examine the case $Z = | \alpha_{\mu e} | = 2.13 \times 10^{-2}$ (democratic). By going through the similar calculation we obtain $X_{0} = 5.90 \times 10^{-2}$ and $Y_{0} = s_{24} = 0.106$, which means $X_{0} / Y_{0} = \sin 2\theta_{14} = 0.557$. Or, $\sin^2 2\theta_{14} = 0.310$, which also passes through the $2 \sigma$ allowed islands. The value of $\sin^2 2\theta_{14}$ of the democratic solution is much closer to the best fit 0.32 of the reactors + Ga data. 

Therefore, we find the appearance-disappearance tension-easing solutions which is consistent with the reactors + Ga combined fit in the leading order version of the non-unitary $(3+1)$ model, for the both $| \alpha_{\mu e} | = 2.54 \times 10^{-2}$ (conservative), and $| \alpha_{\mu e} | = 2.13 \times 10^{-2}$ (democratic) cases. The solutions with the predicted values of $\sin^2 2\theta_{14}$ and $s^2_{24}$ are summarized in the first row of Table~\ref{tab:tension-easing-solution}.

\begin{table}[h!]
\vglue 0.2cm
\begin{center}
\caption{The appearance-disappearance tension easing solutions of the leading-order version of the non-unitary $(3+1)$ model defined in section~\ref{sec:leading-order-model}. In the first column, $A$ denotes the appearance constant, which is read off from the value of $\sin^2 2\theta_{\mu e}$ obtained by the $(3+1)$ model analysis: The first row is for the best fit obtained in ref.~\cite{Dentler:2018sju}, and the second and third show the both ends of the roughly estimated $2 \sigma$ allowed region. 
The second and third columns correspond, respectively, to the ``conservative'' and ``democratic'' bounds on $| \alpha_{\mu e} |$, see section~\ref{sec:diagonal-alpha}. In the fourth column the consistency between our solutions and the (reactors + Ga) and/or the (reactors + solar) combined fits~\cite{Berryman:2021yan} are tabulated with the superscripts $[1]$ and $[2]$, which distinguishes the models with the different $| \alpha_{\mu e} |$ bounds. 
}
\label{tab:tension-easing-solution} 
\vglue 0.2cm
\begin{tabular}{c|c|c|c}
\hline 
$A$ & 
$| \alpha_{\mu e} | = 2.54 \times 10^{-2}$ $^{[1]}$ & 
$| \alpha_{\mu e} | = 2.13 \times 10^{-2}$ $^{[2]}$ & 
Consistent with 
\\
\hline 
\hline 
$6 \times 10^{-3}$ & 
$\sin^2 2\theta_{14} = 0.275$ & 
$\sin^2 2\theta_{14} = 0.310$ & reactors + Ga (2$\sigma$) $^{[1,2]}$ 
\\
%
 & $s^2_{24} = 1.14 \times 10^{-2}$ & 
 $s^2_{24} = 1.13 \times 10^{-2}$ & 
\\
\hline 
$2.7 \times 10^{-3}$ & 
$\sin^2 2\theta_{14} = 0.097$ & 
$\sin^2 2\theta_{14} = 0.113$ & reactors + solar (2$\sigma$) $^{[1,2]}$ 
\\
%
 & $s^2_{24} = 1.08 \times 10^{-2}$ & 
 $s^2_{24} = 1.07 \times 10^{-2}$ & reactors + Ga (3$\sigma$) $^{[2]}$
\\
\hline
$9.3 \times 10^{-3}$ & 
$\sin^2 2\theta_{14} = 0.464$ & 
$\sin^2 2\theta_{14} = 0.515$ & reactors + Ga (3$\sigma$) $^{[1]}$ 
\\
%
 & $s^2_{24} = 1.19 \times 10^{-2}$ & 
 $s^2_{24} = 1.16 \times 10^{-2}$ & \text{no solution} $^{[2]}$
\\
\hline 
\end{tabular}
\end{center}
\vglue -0.2cm 
\end{table}

In view of the appearance and disappearance conditions in eqs.~\eqref{LSND-result} and~\eqref{MINOS-bound}, $\sin 2\theta_{14} | \alpha_{\mu e} |$ must not be too small for the tension-easing mechanism to work. This is the reason why no small $\theta_{14}$ solution, $\sin 2\theta_{14} \ll 1$, is allowed as shown in section~\ref{sec:leading-order-model-1}. But, we learn from Table~\ref{tab:tension-easing-solution} that a modestly small $\theta_{14}$ solution, $\sin^2 2\theta_{14} \sim 0.1$, is allowed for the smallest value of $A$, see subsection~\ref{sec:varying-A}. Overall, our solution prefers large $\theta_{14}$, by which the BEST anomaly, the key element of the reactors + Ga solution, is  ``invited'' to our discussion, in a totally unexpected manner. It is a very interesting feature that the ``tension-easing'' solution serves as a bridge between the two highest confidence level sterile signatures, the LSND-MiniBooNE data and BEST. 

If it turned out, instead, that $\sin^2 2\theta_{14} \sim 0.1$ is preferred by the future studies our model can accommodate it, which predict that we should be in the smallest region of the appearance parameter $A$. 

\subsubsection{Stability with varying $A$}
\label{sec:varying-A}

Let us check the stability of these solutions by varying the appearance constant $A$ within the $2 \sigma$ range $2 \times 10^{-3} \leq A \leq 2 \times 10^{-2}$ (2 DOF), as read off from Fig.~4 in ref.~\cite{Dentler:2018sju}. We can roughly translate the 2 DOF region to quasi-one dimensional 2$\sigma$ allowed region $2.7 \times 10^{-3} \leq A \leq 9.3 \times 10^{-3}$ (1 DOF). 
At the smallest edge of the appearance constant $A = 2.7 \times 10^{-3}$ we obtain $\sin^2 2\theta_{14} = 0.097$ and $s^2_{24} = 1.08 \times 10^{-2}$ for $| \alpha_{\mu e} | = 2.54 \times 10^{-2}$ (conservative). For the democratic case $| \alpha_{\mu e} | = 2.13 \times 10^{-2}$, $\sin^2 2\theta_{14} = 0.113$ and $s^2_{24} = 1.07 \times 10^{-2}$. The both solutions are consistent with the reactor + solar data at $2 \sigma$ CL. A part of the ``democratic'' solution barely overlaps with the $3 \sigma$ region of the reactor + Ga data. 

At the largest edge of $A = 9.3 \times 10^{-3}$ we obtain $\sin^2 2\theta_{14} = 0.464$ and $s^2_{24} = 1.19 \times 10^{-2}$ for $| \alpha_{\mu e} | = 2.54 \times 10^{-2}$ (conservative), and $\sin^2 2\theta_{14} = 0.515$ and $s^2_{24} = 1.16 \times 10^{-2}$ for $| \alpha_{\mu e} | = 2.13 \times 10^{-2}$ (democratic). The ``conservative'' solution is barely consistent with the reactor + Ga data at $3 \sigma$, but ``democratic'' solution has no overlap (just skin touch) with it at $3 \sigma$, as $\sin^2 2\theta_{14}$ is too large. These results are also summarized in Table~\ref{tab:tension-easing-solution}. 

\subsection{Stability check: Bringing back the order unity coefficients} 
\label{sec:1st-order-model} 

To abstract out the leading order model, eq.~\eqref{leading-model}, from the original one given in eqs.~\eqref{P-mue-alpha-1st} and \eqref{P-mumu-alpha-1st}, we have made approximations that the order unity coefficients are set to unity. It includes setting the diagonal $\alpha$ parameters vanish e.g.~in $(1 - 2 \alpha_{ee} - 2 \alpha_{\mu \mu})$, which can be justified because $\alpha_{ee}$ and $\alpha_{\mu \mu}$ are both of the order of $10^{-2}$. But, since we have arrived at the large $\theta_{14}$ solution, the validity of the approximation made by setting $\cos 2\theta_{14} = 1$ and $c^2_{14} = 1$ may look debatable. In our tension-easing solution with $A = 6 \times 10^{-3}$ uncovered in the previous section, $\cos 2\theta_{14} = 0.852$ (0.831) for the conservative (democratic) choices of the $| \alpha_{\mu e} |$ bounds. 

In this section we analyze the ``first-order model'', by which we mean to recover the order unity coefficients in eqs.~\eqref{P-mue-alpha-1st} and \eqref{P-mumu-alpha-1st} which are ignored to construct the leading-order model. We still keep to neglect $s^4_{24} \sin^2 2\theta_{14}$ and the diagonal $\alpha$ parameters. The first-order model can be explicitly written as 
\begin{eqnarray} 
&&
X^2 + 2 \cos 2\theta_{14} X Z = A, 
\nonumber \\
&& 
c^2_{14} c^2_{24} Y^2 
- ( c^2_{24} - s^2_{24} \cos 2\theta_{14} ) X Z = \frac{ B }{4}. 
\label{1st-order-model}
\end{eqnarray}
As in the previous section~\ref{sec:leading-order-model-2}, we denote the zeroth-order solutions, the ones we have obtained by using the leading-order model, as $X_{0}$ and $Y_{0}$. Then, we seek to obtain the first-order corrected solutions with definitions $X = X_{0} + X_{1}$ and $Y = Y_{0} + Y_{1}$ by solving eq.~\eqref{1st-order-model} in the linear approximation in $X_{1}$ and $Y_{1}$. By some simple algebra we obtain 
\begin{eqnarray} 
&&
X_{1} 
= 
\frac{ ( 1 - \cos 2\theta_{14} ) X_{0} Z  }{ \left( X_{0} + \cos 2\theta_{14} Z \right) }, 
\nonumber \\
&& 
Y_{1} 
= 
- \frac{ X_{0} Z }{ 2 Y_{0} } 
+
\frac{1}{ 2 c^2_{14} c^2_{24} Y_{0} } 
\left[
( c^2_{24} - s^2_{24} \cos 2\theta_{14} )  
\frac{ X_{0} Z ( X_{0} +  Z ) } { \left( X_{0} + \cos 2\theta_{14} Z \right)  } 
+ ( 1 - c^2_{14} c^2_{24} ) \frac{ B }{4} 
\right].
\nonumber \\
\label{1st-order-sol}
\end{eqnarray} 

We examine the best-fit $A$ case, our main scenario in the first row in Table~\ref{tab:tension-easing-solution}. Let us calculate the values of $X_{1}$ and $Y_{1}$. In the case of conservative solution ($| \alpha_{\mu e} | = 2.54 \times 10^{-2}$) we obtain $X_{1} = 2.72 \times 10^{-3}$ and $Y_{1} = 5.08 \times 10^{-3}$. Therefore, $X_{1} / X_{0} = 4.84 \times 10^{- 2}$, and $Y_{1} / Y_{0} = 4.75 \times 10^{-2}$. The first order corrections are both less than $5\%$. 
For the democratic solution ($| \alpha_{\mu e} | = 2.13 \times 10^{-2}$), we obtain $X_{1} = 2.76 \times 10^{-3}$ and $Y_{1} = 5.74 \times 10^{-3}$. Therefore, $X_{1} / X_{0} = 4.68 \times 10^{- 2}$, and $Y_{1} / Y_{0} = 5.42 \times 10^{-2}$, showing again $5\%$-$6\%$ level corrections. 

Let us estimate how $\sin 2\theta_{14}$ is affected by including the first order corrections. We obtain by using 
$\sin 2\theta_{14} ^{(0+1)} = ( X_{0} + X_{1} ) / ( Y_{0} + Y_{1} )$ 
\begin{eqnarray} 
&&
\sin 2\theta_{14} ^{(0+1)}
= 0.524 \left( 1 + 0.0009 \right) 
= 0.524 ~~~\text{(conservative)}, 
\nonumber \\
&&
\sin 2\theta_{14} ^{(0+1)}
= 0.557 \left( 1 - 0.0074 \right) 
= 
0.553 ~~~\text{(democratic)}. 
\label{theta14-to-1st}
\end{eqnarray} 
In the conservative case $\sin 2\theta_{14}$ stays the same value with that of the leading-order model, because the difference between $X_{1} / X_{0}$ and $Y_{1} / Y_{0}$ is only $0.1\%$. In the democratic case $\sin 2\theta_{14}$ receive only $- 0.7\%$ correction to the zeroth order value 0.557. Therefore, our leading-order model gives a good approximation to the  first-order corrected model given in eq.~\eqref{1st-order-model}. This is the reason why we present the simpler-to-reproduce, the leading-order model results in Table~\ref{tab:tension-easing-solution}. 

Thus, we have found a few tension-easing solutions as reported in Table~\ref{tab:tension-easing-solution}. But, we should note that our analysis is only for illustrative purpose, and is done with the various simplified settings.  

\subsection{Can our non-unitarity model for easing tension verifiable, or falsifiable?}
\label{sec:falsify-model} 

Here, we briefly discuss how to verify or refute our tension easing solution {\em within} the scope in this section. The characteristic feature of the appearance and disappearance probabilities  in eqs.~\eqref{P-mue-alpha-1st} and~\eqref{P-mumu-alpha-1st} is the presence and absence of {\it CP}- or {\it T}-violating terms, respectively. If the ratio of $\sin (\Delta m^2_{41} L / 2E)$ to $\sin^2 (\Delta m^2_{41} L / 4E)$ terms in $P(\nu_\mu \rightarrow \nu_e)$ is controlled by the ratio of the imaginary to real parts of $\widetilde{\alpha}_{\mu e} = | \alpha_{\mu e} | e^{ i ( \phi_{\mu e} + \phi_{24} ) }$, it is an indication that the tension-easing mechanism due to non-unitarity is working. On general ground, however, {\it CP} violation could occur due to the complex phases of the sterile mixing matrix.\footnote{
{\it CP} or {\it T} odd effect could be produced by the lepton KM phase $\delta$~\cite{Kobayashi:1973fv}. But, this effect would be smaller than the effect we discuss here if we stay on the region where the $\Delta m^2_{41}$-driven sterile oscillation effect dominates over the atmospheric ones. }
Hence, to establish our tension-easing solution, a global fit to all the relevant data is required. 

Conversely, it should be easy to falsify our non-unitary $(3+1)$ model for easing tension. Let us restrict our discussion to the leading-order model as it is reasonably accurate. In Table~\ref{tab:tension-easing-solution} one notices that $\sin^2 2\theta_{14}$ increases when $Z = | \alpha_{\mu e} |$ decrease from the second to third columns. In fact, one can show generally that $\frac{d}{d Z} \sin^2 2\theta_{14} < 0$ by using the expression of $\sin^2 2\theta_{14} = ( X_{0} / Y_{0} )^2$ as a function of $Z$, see eqs.~\eqref{X0-sol} and~\eqref{Y0-sol}. That is, $\sin^2 2\theta_{14}$ is monotonically decreasing function of $Z$. Therefore, when $Z = | \alpha_{\mu e} |$ bound becomes tighter and tighter, $\sin^2 2\theta_{14}$ is monotonically increasing, such that at some point it cannot fit to the reactor + Ga data any more, or even becomes unphysical, $> 1$. 

In section~\ref{sec:Amumu-bound}, we have remarked that the analysis of the MINOS/MINOS+ data under this framework may provide the first signal for consistency of our non-unitarity approach to the solution of the appearance-disappearance tension, or, its failure. 
Coloma {\it et al.} found that with the DUNE near detector with 10 years running, one can achieve the non-unitary $\nu$SM $| \alpha_{\mu e} |$ bound close to 0.01 even with 5\% shape error~\cite{Coloma:2021uhq}. If the similar sensitivity can be reached for the non-unitary $(3+1)$ model's $| \alpha_{\mu e} |$, this would be sufficient to exclude our tension-easing mechanism using non-unitarity. 

A possible different way of falsification of our solution, for example, the reactor + Ga solution. It could happen if the BEST result is excluded experimentally on a firm basis. 

\section{Non-unitary $(3+1)$ model vs. $(3+2)$ or $(3+3)$ model simulations} 
\label{sec:relationship} 

At a theoretical level, the unitary $(3+1+N_{s})$ model and our non-unitary $(3+1)$ model are completely different theories to each other, even though the latter has been started from the former. In the former case, non-unitarity stems from the setting that we restrict ourselves to the measurements of active sector fermions which feel the $\nu$SM interactions. In the latter, if we follow the spirit of the original formulation~\cite{Antusch:2006vwa}, one performs (quantum) integration over the sterile state subspace such that the physical space of the non-unitary theory is automatically that of the active neutrinos in the non-unitary $\nu$SM, or the three active plus one visible sterile state in our case in this paper. However, such integration over the sterile space has not been carried out, and it is beyond the scope of this paper. Then, the distinction between the above two treatments is obscured. The both treatments embody the similar ``wiping out'' procedure of high frequency sterile oscillations. 
We conclude our comparison between the $(3+1+N_{s})$ model and the non-unitary $(3+1)$ model by saying that the advantage of the latter exists in an easier effective treatment of the large $N_{s}$ cases.

\subsection{Apparent puzzle: Successful vs. unsuccessful tension easing}
\label{sec:puzzle} 

In the preceding works people discuss and simulate the unitary, explicit $(3+1+N_{s})$ models typically with $N_{s} = 1$, or 2 visible sterile states, usually denoted as the $(3+2)$ or $(3+3)$ models, to address the problem of the appearance-disappearance tension. See e.g., refs.~\cite{Sorel:2003hf,Maltoni:2007zf,Dasgupta:2021ies,Nelson:2010hz,Giunti:2015mwa,Boser:2019rta,Diaz:2019fwt,Hardin:2022muu,Acero:2022wqg}, and the references cited therein. Apparently, the general consensus in the community is that these explicit model simulations do not show a sufficient tension-easing feature. 

Our non-unitary $(3+1)$ model in this paper, presented through sections~\ref{sec:UV-(3+1)} to~\ref{sec:app-disapp-tension}, is an alternative approach to the above. This model can accommodate one visible sterile state and (in principle) arbitrary number $N_{s}$ of oscillation ``averaged out'' sterile states~\cite{Fong:2016yyh,Fong:2017gke}, see~Appendix~\ref{sec:non-unitary-3+1}. Our analysis in section~\ref{sec:app-disapp-tension}, though under the simplified assumptions and the Okubo's method for constraining the $\alpha$ parameters at a tree level, produced a few tension easing solutions as tabulated in Table~\ref{tab:tension-easing-solution}. 

Why do we succeed in one theory and not in the other? The former is small $N_{s}$ case of the $(3+1+N_{s})$ model, and the latter comes from the same theory. We simply argue that effects of non-unitarity controlled by the $\alpha$ parameters becomes more significant at larger $N_{s}$, the growing $\alpha$ parameter picture at large $N_{s}$ presented in section~\ref{sec:grow-alpha}. 
In principle, one can falsify this interpretation by running the $(3+1+N_{s})$ model simulation with $N_{s} = 10$, for example. However, we suspect that such simulation and deriving the bound on one (or a few) particular parameter(s) are not likely to be practical for $N_{s} = 10$. It requires analysis in the huge parameter space, which renders the required marginalization over the remaining large number of parameters with unknown priors formidable. 

Unfortunately, the above argument cannot be regarded as a solid proof of the validity of our treatment. Most importantly, the rate of increase of $\alpha_{e e}$ and $\alpha_{\mu \mu}$ as $N_{s}$ grows cannot be calculated in a model-independent manner of the sterile sector. For a final, robust answer to the credibility of our analysis results we need a global analysis of the relevant data set. 

\subsection{Indication of enhanced tension-easing feature seen as $N_{s}$ becomes larger?}
\label{sec:indication}

If the $\alpha$ parameters in the non-unitary $(3+1)$ model is small in small $N_{s}$ region, and evolve to a larger value at large $N_{s}$, there might be corresponding changes in the simulation results of the $(3+1+N_{s})$ model. In this sense  comparisons between the $(3+2)$, $(3+3)$, and $(3+4)$ models, if any, are of our keen interests. Though we have examined the literatures, unfortunately, not so many of them can be found apart from those already mentioned, and none for the $(3+4)$ case in our search. 

We quote here the features reported in refs.~\cite{Maltoni:2007zf,Hardin:2022muu}. The authors of ref.~\cite{Maltoni:2007zf} present $\chi^2_{(3+2)} - \chi^2_{(3+3)}$ per added degrees of freedom as $1.7/4$ (see Table II), and the CL 20\% at which $(3+2)$ is accepted with respect to $(3+3)$, as derived for four additional parameters in the $(3+3)$ model. 
More importantly, the authors of ref.~\cite{Hardin:2022muu} report the results of parameter goodness (PG) of fit test for the $(3+1)$, $(3+2)$, and $(3+3)$ models with $\chi^2_{ \text{PG} } = \chi^2_{ \text{glob} } - (\chi^2_{ \text{app} } - \chi^2_{ \text{dis} })$, an indicator for the tension between the two subsets of data. The $p$ values reported are $7.9 \times 10^{-7}$, $9.3 \times 10^{-8}$, and $4.9 \times 10^{-5}$, respectively, for the $(3+1)$, $(3+2)$, and $(3+3)$ models. Though small in numbers of the $p$ values, there exists a marked difference between the $(3+2)$ and $(3+3)$ models, the most significant change in the $p$ value in this analysis. 
The above quoted results in refs.~\cite{Maltoni:2007zf,Hardin:2022muu} may be interpreted as an indication that adding the second averaged-out sterile states may have enhanced the tension easing. This feature is in harmony with our above picture of more enhanced tension-easing at larger $N_{s}$. 

\section{Concluding remarks}
\label{sec:conclusion}

In this paper we have addressed so called the problem of ``appearance-disappearance tension'' between the LSND-MiniBooNE measurement of $P(\nu_\mu \rightarrow \nu_e)$, and the MINOS (and others) measurement of $P(\nu_\mu \rightarrow \nu_\mu)$, in its sterile neutrino interpretation. We have assumed the basic framework of $(3+1)$ model to accommodate the single (almost) sterile state into the $\nu$SM. To embody our understanding of non-unitarity as the most natural interpretation of the tension, we have constructed the non-unitary $(3+1)$ model and presented an illustrative analysis to demonstrate that the idea works, under the various simplifications including ignoring the $\nu$SM oscillations, the matter effect,  and the sterile parameters' CP phase effects. 

We would like to emphasize that our approach to resolving tension for eV-scale sterile states via non-unitarity, though truly novel (we believe) at the level of idea, is still in its infancy. It means that many things remain to be understood, including the required analysis machinery as well as implications of the scenario. 
Possibly the most important implication would be the ``transported into solution'' large neutrino-antineutrino asymmetry. Our system, a priori, does not involve a seed for neutrino-antineutrino asymmetry in its definition, in particular, in the disappearance channel probabilities, eqs.~\eqref{P-ee} and~\eqref{P-mumu-alpha-1st}. But, it displays the large neutrino-channel anomaly a posteriori by inviting the BEST anomaly. On the other hand, apparently, there is no comparably significant anomaly in the antineutrino channels. See the next two sections~\ref{sec:existence} and~\ref{sec:future}. 

It might be that such asymmetry is only characteristic to the sterile-related sector, not in the whole $\nu$SM, although we do not know if such system allows a suitable formulation. We certainly do hope that these implication discussions are not to be masked by the technical issues involved in our {\em present} analysis. 

\subsection{Existence of the tension-easing solution} 
\label{sec:existence}

Do we find the solution to the appearance-disappearance tension in our non-unitary $(3+1)$ model? The answer is {\em Yes} in the sense that the analysis presented in section~\ref{sec:app-disapp-tension} may be regarded as an illustrative, semi-quantitative one, a precursor to the complete analysis with the $\alpha$ parameters constrained by a global analysis of the relevant data. Though the latter must be doable, such analysis is beyond the scope of this paper. 

Now, we would like to highlight the particular solution with the unique character, from all the solutions given in Table~\ref{tab:tension-easing-solution}. At the best-fit value of the appearance constant $A$~\cite{Dentler:2018sju}, we have obtained the unique {\em ``robust and clean''} solution which predicts the value of $\sin^2 2\theta_{14} = 0.275$ and $\sin^2 2\theta_{14} = 0.310$ corresponding, respectively, to the conservative and democratic choices of the $| \alpha_{\mu e} |$ bound, see section~\ref{sec:diagonal-alpha}. In the both cases, the solutions are inside $2 \sigma$ CL allowed contours of the reactor + Ga data, as analyzed and presented in ref.~\cite{Berryman:2021yan}. By ``robust'' we mean the same (reactor + Ga) solution is obtained for the both cases of conservative and democratic $| \alpha_{\mu e} |$ bounds, and exists in most of the allowed region of the $A$ parameter. By ``clean'' we mean that no other solution is allowed except for this one in any one of the examined two $| \alpha_{\mu e} |$ bounds.\footnote{
We note that the small $\sin \theta_{14} \sim 0.1$ solution does exist albeit not the ``robust and clean'' type. See Table~\ref{tab:tension-easing-solution}. }
We note that this solution is largely driven by the $^{51}$Cr source experiment BEST~\cite{Barinov:2021asz,Barinov:2022wfh}, which sees about $\sim$20\% deficit of $\nu_e$. Thus, our tension easing solution for eV-scale sterile(s) serves as a ``bridge'' between the two independent high-CL phenomena, the LSND-MiniBooNE anomaly and BEST. 

A  cautionary remark must be added about our current estimates of the $\alpha$ parameter bounds used in the analysis in section~\ref{sec:app-disapp-tension}. Lacking the global analysis, we have carried out a tree level estimate of the $\alpha$ parameters by using the Okubo's method which allows us to express the $\alpha$ parameters by the mixing angles and phases of the larger, unitary theory. See section~\ref{sec:okubo-brief} and Appendix~\ref{sec:Okubo-construction}. Despite that our numbers are at best the plausible estimates, they are the reasonable ones obtained by the available best method, to our knowledge. 

Here is a remark on solar neutrino bound on $\alpha_{ee}$: One can show that the bound obtained by using the Okubo's method excludes the BEST result, which may be a part of ``solar-neutrino$-$BEST'' tension~\cite{Berryman:2021yan,Giunti:2022btk}. This is comparison of $\nu_{e}$ both in matter and in vacuum. 
Independently, there is an indication that $\Delta m^2_{21}$ measured by $\bar{\nu}_{e}$ in vacuum (KamLAND~\cite{KamLAND:2013rgu}) and its solar neutrino measurement differ by about $1.5 \sigma$~\cite{Super-Kamiokande:2023jbt}. Recently the KamLAND results is conformed by SNO+~\cite{SNO:2025chx} and JUNO~\cite{JUNO:2025gmd} with the similar and 1.6 times improved accuracies. If these results signal an exotic character of the matter effect in the Sun, it could affect the interpretation of the solar-neutrino$-$BEST tension. 

\subsection{Possible neutrino-antineutrino asymmetry and future perspectives}
\label{sec:future}

We have started this paper by mentioning the two major obstacles against establishing the existence of the eV-scale sterile neutrino(s). One is the problem of tension for which we have proposed our own solution by introducing low-scale non-unitarity. The other problem, most probably the severer one, is the tension with cosmology. It appears that the stringent cosmological constraints on sterile(s), see e.g., ref.~\cite{Hagstotz:2020ukm}, makes inevitable to introduce a new ingredient into the standard $\Lambda$CDM, see e.g.,~\cite{Perivolaropoulos:2021jda}. Self-interactions among sterile states looks a good candidate for this purpose~\cite{Dasgupta:2021ies,Hannestad:2013ana,Dasgupta:2013zpn,Archidiacono:2014nda,Archidiacono:2015oma}, as mentioned in section~\ref{sec:introduction}. 

We notice that the above candidate solutions for these major issues on sterile(s) jointly present a radically different view of matter from what we know now. An example would be a feebly self-interacting sterile matter of the large~$N_{s}$ ``background'' sterile states, though we do not know if such view can bear resemblance to physical reality. Fortunately, we will know quite soon what the ongoing and upcoming experiments~\cite{Ajimura:2017fld,MicroBooNE:2015bmn} will tell us about the questions on eV-scale sterile. 

In section~\ref{sec:introduction}, we have mentioned that the recent results of the several sterile-related search experiments do not appear to converge. If the reactor antineutrino anomaly (RAA)~\cite{Mention:2011rk,Mueller:2011nm,Huber:2011wv} is largely cured by the beta decay electron energy spectrum measurement by Kopeikin {\it et al.}~\cite{Kopeikin:2021ugh}, see, e.g., ref.~\cite{Giunti:2021kab}, we observe a large neutrino-antineutrino asymmetry: A 20\%-level large deficit in the neutrino channel~\cite{Barinov:2021asz,Barinov:2022wfh}, and much less anomaly in the antineutrino channel. 
On the other hand, the precision tritium beta decay measurement KATRIN~\cite{KATRIN:2024cdt,KATRIN:2025lph} excluded (95\% CL) most of the region favored by the BEST result. 
If all these experimental results are correct what would be a unifying picture? It appears to the author that the only solution is a large unknown anomalous effect 
in the neutrino channel (BEST), and no (or small) anomaly in the antineutrino channel (RAA and KATRIN). But, it implies violation of {\it CPT} in vacuum, the fundamental symmetry of quantum field theory~\cite{Itzykson:1980rh}. We close only with questions: Is the asymmetry limited in the sterile-related sector? Is there any sensible formulation of such system? 

Are there ways to settle this issue experimentally? If we suspect that the radioactive source measurement can somehow hide problems, several methods for clarification are proposed. 
(1) Gavrin {\it et al.}~propose the BEST-2 experiment using $^{58}$Co neutrino source as a cross check of the BEST result and measurement of the relevant $\Delta m^2$~\cite{Gavrin:2025ctr}. 
(2) A scintillator experiment with cerium-doped gadolinium aluminum gallium garnet (Ce:GAGG) is proposed~\cite{Huber:2022osv} for the simultaneous two-channel measurement of gallium capture events and neutrino electron scattering events, whose latter serves for an in-situ source strength measurement. 
(3) For possible direct relevance to the issue of large neutrino-antineutrino asymmetry, which may be related to {\it CPT} violation, the Cerium 144 $\bar{\nu}_{e}$ source experiment which was proposed sometime ago~\cite{Gaffiot:2014aka,Gando:2013zoa} should bear a renewed attention.

\begin{acknowledgments} 

The author thanks Enrique~Fernandez-Martinez for illuminating, critical discussions on non-unitarity and the bounds on it. Our communications span a long period of time started from H.M.'s stay in Instituto F\'{\i}sica Te\'{o}rica in Madrid in 2018. 
He expresses gratitude to the anonymous authors for the numerous comments which are proved to be valuable to sharpen up the scientific discussions in this manuscript. The informative correspondences with Thierry Lasserre about the Cerium source experiment and the discussions with Kimihiro Okumura on the SK atmospheric neutrino analyses were essential to make the author's understanding clearer, bringing this manuscript into the level as it is now. 

\end{acknowledgments}

\appendix 

\section{Partial-unitarity correlation} 
\label{sec:partial-unitarity} 

In section~\ref{sec:UV-natural}, we are motivated to our non-unitarity approach by saying that ``the disappearance measurements do not observe sufficient event number depletion expected by unitarity''. Obviously a question may arise about if it makes sense because unitarity in the $(3+1)$ model should involve $\nu_{\tau}$ and $\nu_{S}$. Here is some explanation about what it actually means. 

The expressions of $P(\nu_\mu \rightarrow \nu_e)$ and $P(\nu_\mu \rightarrow \nu_\mu)$ in vacuum in our simplified $(3+1)$ model can be obtained by setting the $\alpha$ parameters vanish in eqs.~\eqref{P-mue-alpha-1st} and \eqref{P-mumu-alpha-1st}, respectively. The probabilities in the remaining channels in the $\nu_\mu$ row are given by 
\begin{eqnarray}
&& 
P(\nu_\mu \rightarrow \nu_\tau) = 
s^2_{34} c^4_{14} \sin^2 2\theta_{24} 
\sin^2 \left( \frac{\Delta m_{41}^2 L}{4E} \right), 
\nonumber\\
&& 
P(\nu_\mu \rightarrow \nu_S) 
= 
c^2_{34} c^4_{14} \sin^2 2\theta_{24} 
\sin^2 \left( \frac{\Delta m_{41}^2 L}{4E} \right), 
\label{P-mutau-muS-vac} 
\end{eqnarray}
with which one can prove unitarity, 
\begin{eqnarray}
P(\nu_\mu \rightarrow \nu_e) + P(\nu_\mu \rightarrow \nu_\mu) - 1 
&=&
- \left[ P(\nu_\mu \rightarrow \nu_\tau) + P(\nu_\mu \rightarrow \nu_S) \right] 
\nonumber\\
&=&
- c^4_{14} \sin^2 2\theta_{24} 
\sin^2 \left( \frac{\Delta m_{41}^2 L}{4E} \right). 
\label{unitarity-mu-row} 
\end{eqnarray}
In the last line in eq.~\eqref{unitarity-mu-row} we give the explicit expression, anticipating ``partial unitarity'' discussion given below. If it were vanishing, it implies the $\nu_{e} - \nu_{\mu}$ two channel unitarity, but of course, it is not the case. Nonetheless, one notices that $\nu_{e} - \nu_{\mu}$ sub-sector is special because, for example, if $s_{34} = 0$, $\nu_{\tau}$ decouples and $P(\nu_\mu \rightarrow \nu_\tau)$ vanishes, see eq.~\eqref{P-mutau-muS-vac} and eq.~\eqref{U-def} for the $U$ matrix. 

We introduce some simple notations 
$P(\nu_\mu \rightarrow \nu_e) = \mathcal{A} \sin^2 \left( \Delta m_{41}^2 L / 4E  \right)$, 
$1 - P(\nu_\mu \rightarrow \nu_\mu) = \mathcal{D} \sin^2 \left( \Delta m_{41}^2 L / 4E  \right)$, where  
$\mathcal{A} \equiv s^2_{24} \sin^2 2\theta_{14}$ and 
$\mathcal{D} \equiv c^2_{14} \sin^2 2\theta_{24}$. 
We have simplified $\mathcal{D}$ by ignoring the $s^4_{24}$ term~$\lsim 10^{-4}$. We also define $\mathcal{U} \equiv c^4_{c4} \sin^2 2\theta_{24}$ in the right-hand side of eq.~\eqref{unitarity-mu-row}. 
Let us consider the simultaneous variations of $s^2_{24}$ and $s^2_{14}$ under which $\mathcal{U}$ is invariant, 
\begin{eqnarray}
&&
\frac{ d }{ d \Xi } \mathcal{U} 
\equiv 
\left( \sin^2 2\theta_{24} \frac{ \partial  }{ \partial s^2_{24} } 
+ 2 c^2_{14} \cos 2\theta_{24} \frac{ \partial }{ \partial s^2_{14} } \right) 
\mathcal{U} 
= 0. 
\label{U-preserving}
\end{eqnarray}
Then, along the $\Xi$ direction one can show that $\mathcal{A}$ and $\mathcal{D}$ vary as 
\begin{eqnarray}
\frac{ d }{ d \Xi } \mathcal{A} 
&=&
\sin^2 2\theta_{24} \sin^2 2\theta_{14} 
+ 
8 s^2_{24} c^2_{14} \cos 2\theta_{24} \cos 2\theta_{14} > 0,  
\nonumber \\
\frac{ d }{ d \Xi } \mathcal{D}  
&=&
2 c^2_{14} \cos 2\theta_{24}\sin^2 2\theta_{24} > 0,  
\label{A-D-variation}
\end{eqnarray}
where we have assumed that $0 < \theta_{24} < \pi/4$ and $0 < \theta_{14} < \pi/4$. 

The meaning of this exercise is as follows: 
Under the $\mathcal{U}$ preserving variations of $s^2_{24}$ and $s^2_{14}$, the right-hand side of eq.~\eqref{unitarity-mu-row} stays constant. Therefore, 
$P(\nu_\mu \rightarrow \nu_e) + P(\nu_\mu \rightarrow \nu_\mu) = \mathcal{O} [1]$, an order unity {\em constant} under the variations. 
This is not a precise $\nu_{e} - \nu_{\mu}$ sub-sector unitarity, but guarantees that the similar correlation between $P(\nu_\mu \rightarrow \nu_e)$ and $P(\nu_\mu \rightarrow \nu_\mu)$ is functional: $P(\nu_\mu \rightarrow \nu_e) = 1 - P(\nu_\mu \rightarrow \nu_\mu)$ - (small constant) under the $\mathcal{U}$ preserving variations. 
When $P(\nu_\mu \rightarrow \nu_e)$ becomes larger, $1 - P(\nu_\mu \rightarrow \nu_\mu)$ becomes larger at the same time, which causes event depletion in the disappearance channel. This structure may be called as ``partial unitarity'', or partial-unitarity correlation between $P(\nu_\mu \rightarrow \nu_e)$ and $P(\nu_\mu \rightarrow \nu_\mu)$. 

Our message delivered in section~\ref{sec:UV-natural}, which is repeated at the beginning of this Appendix, sounds like that we have assumed the $\nu_{e} - \nu_{\mu}$ sub-sector unitarity. We did not, but the statement itself is valid in the above sense.

\section{Construction of the non-unitarity $(3+1)$ model} 
\label{sec:non-unitary-3+1} 

In this Appendix we briefly describe how the non-unitary $(3+1)$ model can be constructed starting from the system of three active $\nu$SM neutrinos, one ``visible''  eV-scale dominantly sterile neutrino, and the averaged out $N_{s}$ sterile states. Our presentation essentially follows that in refs.~\cite{Fong:2016yyh,Fong:2017gke} which treat the non-unitarity $\nu$SM, but it is easy to convert the formulation to the non-unitarity $(3+1)$ model. 

\subsection{The $(3+1)$ model with non-unitarity}
\label{sec:3+1-UV}

In section~\ref{sec:implementing-UV}, we have introduced the $(4+N_{s}) \times (4+N_{s})$ unitary mixing matrix ${\bf U}$ defined by eq.~\eqref{bfU-(3+N)} which connect the mass eigenstates $\nu_{k}$ ($k=1,2,3,4,J$) to the flavor eigenstate of the $(3+1+N_{s})$ model, where the label $J$ runs over the $N_{s}$ sterile neutrino subspace. Under certain kinematic conditions, such as $\Delta m^2_{4 i} \simeq (1-10)$~eV$^2$ ($i=1,2,3$) and $\Delta m^2_{J 4} \sim 100$~eV$^2$, one can show that the $N_{s}$ sterile - active and sterile - sterile oscillations are averaged out. Then, the system can be interpreted as the one composed of the three active plus one eV-scale sterile neutrinos with non-unitarity~\cite{Fong:2016yyh,Fong:2017gke}. We have investigated the problem of how the $S$ matrix and the probability should be calculated in theories with non-unitarity. 

Using this modified framework of the one given in ref.~\cite{Fong:2016yyh}, we obtain the expression of the oscillation probability measured at distance $x$ in vacuum. In the appearance channel $\alpha \neq \beta$ the probability is given by 
\begin{eqnarray}
P(\nu_\beta \rightarrow \nu_\alpha) 
&=& 
\mathcal{C}_{\alpha \beta} + 
\biggl | \sum_{j=1}^{4} N_{\alpha j} N^{*}_{\beta j} \biggr |^2 
- 4 \sum_{ j < k \leq 4 } 
\mbox{Re} \left( N_{\alpha j} N_{\beta j}^* N_{\alpha k}^* N_{\beta k} \right) 
\sin^2 \frac{ \Delta m^2_{kj} x  }{ 4E } 
\nonumber\\
&-&
2 \sum_{ j < k \leq 4 } 
\mbox{Im} \left( N_{\alpha j} N_{\beta j}^* N_{\alpha k}^* N_{\beta k} \right) 
\sin \frac{ \Delta m^2_{kj} x  }{ 2E }, 
\label{P-beta-alpha-ave-vac}
\end{eqnarray}
and in the disappearance channel by 
\begin{eqnarray}
P(\nu_\alpha \rightarrow \nu_\alpha) = 
\mathcal{C}_{\alpha \alpha } + 
\biggl( \sum_{j=1}^{4} \vert N_{\alpha j} \vert^2 \biggr)^2 
- 4 \sum_{ j < k \leq 4 }  \vert N_{\alpha j} \vert^2 \vert N_{\alpha k} \vert^2 
\sin^2 \frac{ \Delta m^2_{kj} x  }{ 4E }. 
\label{P-alpha-alpha-ave-vac}
\end{eqnarray}
Simplification in the probability formulas in eqs.~\eqref{P-beta-alpha-ave-vac} and~\eqref{P-alpha-alpha-ave-vac}, in particular, the absence of the sterile frequencies $\Delta m^2_{J 4} x / 4 E$ or $\Delta m^2_{J 4} x / 2 E$ occurs because they are averaged out to constant values, or become negligibly small due to suppression of the energy denominator. We have shown that this mechanism works in vacuum~\cite{Fong:2016yyh} as well as in matter~\cite{Fong:2017gke}. 

The expressions of the probability formulas in eqs.~\eqref{P-beta-alpha-ave-vac} and~\eqref{P-alpha-alpha-ave-vac} are akin to the usual vacuum probability formulas in the $\nu$SM, at first glance just replacing the $U$ matrix by the non-unitary $N$ matrix. However, there are crucial differences in the first two constant terms. 
In eqs.~\eqref{P-beta-alpha-ave-vac} and ~\eqref{P-alpha-alpha-ave-vac}, $\mathcal{C}_{\alpha \beta}$ and $\mathcal{C}_{\alpha \alpha }$ denote the probability leaking terms~\cite{Fong:2016yyh,Fong:2017gke}
\begin{eqnarray} 
\mathcal{C}_{\alpha \beta} \equiv 
\sum_{J=5}^{5+N_{s}}
\vert W_{\alpha J} \vert^2 \vert W_{\beta J} \vert^2, 
\hspace{10mm}
\mathcal{C}_{\alpha \alpha } \equiv 
\sum_{J=5}^{5+N_{s}} 
\vert W_{\alpha J} \vert^4.
\label{Cab-Caa}
\end{eqnarray}
Interestingly, the forms of $\mathcal{C}_{\alpha \beta}$ and $\mathcal{C}_{\alpha \alpha }$ remains the same in the matter environments~\cite{Fong:2017gke}. They exist because the probability leaks from the $4 \times 4$ (3 active$+\nu_{S}$) state space to the decohered $N_{s} \times N_{s}$ background sterile space. 

The upper and lower bounds on the probability leaking terms are derived for the non-unitary $\nu$SM. It is a simple task to re-derive them in our non-unitary $(3+1)$ model context. If we denote the right-hand side of eq.~\eqref{Cauchy-Schwartz-def} as $\text{RHS}_{(5.4)}$, the bounds read: $( 1/N_{s} ) \text{RHS}_{(5.4)} \leq \mathcal{C}_{\alpha \beta} \leq \text{RHS}_{(5.4)}$. For $\mathcal{C}_{\alpha \alpha}$ we take $\beta = \alpha$. 
For more about interpretation of the probability leaking terms, see refs.~\cite{Fong:2016yyh,Fong:2017gke}. 

Another new feature exists in the second terms in eqs.~\eqref{P-beta-alpha-ave-vac} and~\eqref{P-alpha-alpha-ave-vac}, the ``mis-normalization'' terms. In unitary theory, it vanishes in the appearance channel and it is unity in the disappearance channel. 

In the $(3+1+N_{s})$ model the whole theory is unitary, ${\bf U} {\bf U}^\dagger = 1_{(4+N_{s}) \times (4+N_{s})}$. It leads to 
\begin{eqnarray} 
\alpha + \alpha^{\dagger} - \alpha \alpha^{\dagger} 
= W W^{\dagger}. 
\end{eqnarray}
Therefore, $\alpha \sim |W|^2$~\cite{Fong:2017gke}. Then, the probability leaking terms $\mathcal{C}_{\alpha \beta}$ and $\mathcal{C}_{\alpha \alpha}$, which are of order $|W|^4$, are of order $\alpha^2$ in terms of the $\alpha$ parameters. 
Notice that the degree of freedom of the $4 \times 4$ $\alpha$ matrix is 16, and of $W$ is $8 N_{s}$. Therefore, when $N_{s}$ becomes large the relation between the $\alpha$ parameters and the sterile-active mixing angles becomes less and less tight. In the $(3+N_{s})$ model, the similar discussion goes through.

\subsection{The probabilities $P(\nu_\mu \rightarrow \nu_e)$ and $P(\nu_\mu \rightarrow \nu_\mu)$ }
\label{sec:P-mue-mumu}

For use in our analysis in section~\ref{sec:app-disapp-tension}, we present the oscillation probabilities $P(\nu_\mu \rightarrow \nu_e)$ and $P(\nu_\mu \rightarrow \nu_\mu)$ in our non-unitary $(3+1)$ model in vacuum. We simply give here the expressions in the neutrino channel, but the one in anti-neutrino channel can be obtained by taking complex conjugate of the CP phase related quantities of the form $e^{\pm i \phi}$. We use the $\alpha$ parametrization of the $N$ matrix, $N = (1 - \alpha) U$, whose matrix elements are easily calculable with $U$ in eq.~\eqref{U-def} and the $\alpha$ matrix elements in eq.~\eqref{alpha-def}. 

We leave the probability leaking term~\cite{Fong:2016yyh,Fong:2017gke}, $\mathcal{C}_{\mu e}$ and $\mathcal{C}_{\mu \mu}$, as they are, but they cannot be uniquely specified without making further assumptions. $P(\nu_\mu \rightarrow \nu_e)$ and $P(\nu_\mu \rightarrow \nu_\mu)$ are given by 
\begin{eqnarray}
&&
P(\nu_\mu \rightarrow \nu_e) 
= 
\mathcal{C}_{e \mu} 
+ ( 1 - \alpha_{e e} )^2 | \alpha_{\mu e} |^2 
\nonumber \\
&+&
( 1 - \alpha_{e e} )^2 \sin 2\theta_{14} 
\biggl\{
( 1 - \alpha_{\mu \mu} )^2 s^2_{24} \sin 2\theta_{14} 
- | \widetilde{\alpha}_{\mu e}  |^2 \sin 2\theta_{14} 
+ 2 ( 1 - \alpha_{\mu \mu} ) s_{24} 
\cos 2\theta_{14} \mbox{Re} \left( \widetilde{\alpha}_{\mu e} \right)  
\biggr\} 
\nonumber \\
&\times& 
\sin^2 \frac{\Delta m_{41}^2 L}{4E} 
\nonumber \\
&-& 
( 1 - \alpha_{e e} )^2 ( 1 - \alpha_{\mu \mu} ) s_{24} 
\sin 2\theta_{14} 
\mbox{Im} \left( \widetilde{\alpha}_{\mu e} \right) 
\sin \frac{\Delta m_{41}^2 L}{2E}. 
\label{P-mue-full} 
\end{eqnarray} 
\begin{eqnarray} 
&& 
P(\nu_\mu \rightarrow \nu_\mu) 
= 
\mathcal{C}_{\mu \mu} 
+ 
\left\{ ( 1 - \alpha_{\mu \mu} )^2 + | \widetilde{\alpha}_{\mu e} |^2 \right\}^2 
\nonumber\\
&-& 
4 \biggl\{
( 1 - \alpha_{\mu \mu} )^2 s^2_{24} c^2_{14} 
+ | \alpha_{\mu e} |^2 s^2_{14} 
- ( 1 - \alpha_{\mu \mu} ) \mbox{Re} \left( \widetilde{\alpha}_{\mu e} \right) 
s_{24} \sin 2\theta_{14} 
\biggr\} 
\nonumber\\
&\times&
\biggl\{ 
( 1 - \alpha_{\mu \mu} )^2 ( c^2_{24} + s^2_{24} s^2_{14} ) 
+ | \widetilde{\alpha}_{\mu e} |^2 c^2_{14} 
+ ( 1 - \alpha_{\mu \mu} ) \mbox{Re} \left( \widetilde{\alpha}_{\mu e} \right) 
s_{24} \sin 2\theta_{14} 
\biggr\} 
\sin^2 \frac{ \Delta m^2_{41} x }{ 4E }. 
\nonumber\\
\label{P-mumu-full} 
\end{eqnarray} 

\section{Bounds on the sterile mixing angles $s^2_{14}$ and $s^2_{24}$} 
\label{sec:sterile-mixing-bound}

As indicated in sections~\ref{sec:oscillation-P} and~\ref{sec:s14-24-bound}, it is never straightforward to derive the bound on the sterile mixing angles $s^2_{14}$ and $s^2_{24}$ in our non-unitary $(3+1)$ model setting. The existing bounds on the $(3+1)$ model parameters are derived within the framework of the {\em unitary} $(3+1)$ model, and non-unitarity makes the job much more complicated by the coexisting $\alpha$ parameters which are again different from the existing ones for the non-unitary $\nu$SM. But, let us try to find the way we circumvent the difficulties. 

\subsection{Constraints on the sterile mixing angles: $s^2_{14}$} 
\label{sec:s14-bound} 

As we take the appearance events corresponding to the eV-scale sterile neutrino for granted, $s_{14}$ should not vanish. Otherwise the whole probability $P(\nu_\mu \rightarrow \nu_e)$ vanishes, apart from the constant terms, even after including non-unitarity, see eq.~\eqref{P-mue-alpha-1st} or eq.~\eqref{P-mue-full}. However, the question of whether $s_{14}$ is non-vanishing or not, and which value $s_{14}$ takes if non-zero, does not appear to have an affirmative answer experimentally at this moment. 

A promising way of accessing to the value of $s_{14}$ is to carry out the SBL reactor neutrino experiments~\cite{Danilov:2024fwi,NEOS:2016wee,PROSPECT:2020sxr,STEREO:2019ztb,Serebrov:2020kmd}. In the $(3+1)$ model extended with non-unitarity the $\nu_{e}$ (and $\bar{\nu}_{e}$) survival probability in vacuum is given by 
\begin{eqnarray} 
&&
P(\nu_e \rightarrow \nu_e) 
= P(\bar{\nu}_{e} \rightarrow \bar{\nu}_{e}) 
= 
( 1 - \alpha_{ee} )^4 
\biggl( 1 -  \sin^2 2\theta_{14} \sin^2 \frac{ \Delta m^2_{41} x  }{ 4E } \biggr). 
\label{P-ee}
\end{eqnarray}
Then, the question is how we can determine $s^2_{14}$ under the coexisting unknown parameter $\alpha_{ee}$. Our answer is to make a normalization-free analysis with the survival probability in eq.~\eqref{P-ee}, which would reduce the analysis to the one of the unitary $(3+1)$ model, allowing us to determine $\sin^2 2\theta_{14} = 4 | U_{e4} |^2 ( 1 - | U_{e4} |^2 )$.\footnote{
It is customary to use $\sin^2 2\theta$ and $\Delta m^2$ using the ``two-flavor'' fit in analyzing the results of SBL reactor experiments. However, we translate the notations for clarity (and brevity) to the ones of the corresponding quantities in the $(3+1)$ model defined in section~\ref{sec:3+1-vac}. }
There exist global analyses of these experiments using the $(3+1)$ model, see e.g., refs.~\cite{Dentler:2018sju,Berryman:2021yan,Hardin:2022muu}. Among them, 
Berryman {\it et al.}~\cite{Berryman:2021yan} declare that their analysis is based on relative measurements, and therefore, we consult to this reference to know the reasonable values of $s^2_{14}$ to refer in our analysis. 

We find, quite surprisingly, that the best fit values of $\sin^2 2\theta_{14}$ and $\Delta m^2_{41}$ vary a lot from one experiment to another. For example, the best fit for $( \sin^2 2\theta_{14}, \Delta m^2_{41} )$ varies from $(0.014, 1.3~\mbox{eV}^2)$ of DANSS to $(0.63, 8.95~\mbox{eV}^2)$ of STEREO, a big change of a factor of 45 in $\sin^2 2\theta_{14}$. The best fit for all the SBL reactor experiments used in ref.~\cite{Berryman:2021yan} is located at $( \sin^2 2\theta_{14}, \Delta m^2_{41} ) = (0.26, 8.86~\mbox{eV}^2)$, with $1.1 \sigma$ ($2.2 \sigma$ if Wilks' theorem holds) significance of observing the sterile. 
Furthermore, these minima are unstable to inclusion of the data of the solar neutrino observation or the Ga source experiments. For the reactors + solar: $( \sin^2 2\theta_{14}, \Delta m^2_{41} ) = (0.014, 1.30~\mbox{eV}^2)$, and for the reactors + Ga: $(0.32, 8.86~\mbox{eV}^2)$. See Table~1 of ref.~\cite{Berryman:2021yan} and the description in the text for more details. 
Another notable feature is that while the combinations of data, the reactors~vs.~solar, and the reactors~vs.~Ga, are both compatible to each other, there exists strong tension between the solar and the Ga data with $p$ values of order $10^{-4} - 10^{-3}$~\cite{Berryman:2021yan}, see Table~5 and the description in the text for more details. The similar observation is made in ref.~\cite{Giunti:2022btk}. 

Given the above contrived features of the experimental data on $\sin^2 2\theta_{14}$ including the question of whether it is nonzero or not, we lack a reasonable way of uniquely identifying the value of $s_{14}$ for our analysis. Therefore, we rely on Fig.~7 in ref.~\cite{Berryman:2021yan} which present the confidence regions at $1\sigma, 2\sigma, 3\sigma$ for the reactors + solar and the reactors + Ga data. We pick up, arbitrarily, the three values for the candidate points to refer in our analysis, as given in eq.~\eqref{theta-14-input}: $\sin^2 2\theta_{14} = 0.1$, $\sin^2 2\theta_{14} = 0.014$, $\sin^2 2\theta_{14} = 0.32$, roughly representing in order, the high and low $\Delta m^2$ regions of the reactor + solar data, and the reactors + Ga best fit. 

To convert these values of $\sin^2 2\theta_{14} = 4 (1 - s^2_{14} ) s^2_{14}$ into the ones of $s^2_{14}$ we assume that we always pick the smaller solution. For example, for the above second solution, we obtain the two solutions, $s^2_{14} = 3.51 \times 10^{-3}$ and $s^2_{14} = 0.996$, but we choose the former. The other two choices of $s^2_{14}$ are, therefore, given by $s^2_{14} = 0.0257$ and $s^2_{14} = 0.0877$ for the first and the third choices in eq.~\eqref{theta-14-input}, respectively. 

\subsection{Constraints on the sterile mixing angles: $s^2_{24}$} 
\label{sec:s24-bound} 

MINOS and MINOS+ use the charged-current (CC) $\nu_{\mu}$ disappearance measurements to constrain $s^2_{24}$~\cite{MINOS:2017cae}. While the neutral-current (NC) reactions are also analyzed, it appears that the constraints on $s^2_{24}$ and $\Delta m^2_{41}$ dominantly come from the CC reaction channels. They employ the near-far two-detector fit for a higher sensitivity to sterile oscillation compared to the far-over-near ratio method used in the previous analysis~\cite{MINOS:2016viw}. 

Remarkably, the analysis result reveals a very interesting feature. While we naively expect sensitivity improvement dominantly in high $\Delta m^2_{41}$ region with the MINOS/MINOS+ setting, the better sensitivity is obtained, in fact, more or less uniformly in the wide range of $\Delta m^2_{41}$, $10^{-2}~\mbox{eV}^2 \lsim \Delta m^2_{41} \lsim 100~\mbox{eV}^2$, see Fig.~4 in ref.~\cite{MINOS:2017cae}. To our understanding, this owes to the power of the two-detector setting which has sensitivity to different phases of the sterile oscillations depending upon $\Delta m^2_{41}$. 
Focusing on region of our interest, $1~\mbox{eV}^2 \lsim \Delta m^2_{41} \lsim 100~\mbox{eV}^2$, they state~\cite{MINOS:2017cae} that ``oscillations occur in the ND along with rapid oscillations averaging in the FD''. 
The bound they obtained is $s^2_{24} \lsim 10^{-2}$ at 90\% CL in region $1~\mbox{eV}^2 \lsim \Delta m^2_{41} \lsim 10~\mbox{eV}^2$, see Fig.~3 in ref.~\cite{MINOS:2017cae}. 

Now we must address here the question of whether the MINOS/MINOS+ bound on $s^2_{24}$ holds also in our setting in which the $\alpha$ parameter dependent terms exist. Let us ignore, momentarily, the $\widetilde{\alpha}_{\mu e}$ term. Then, the effect of the $\alpha$ parameters is through the $( 1 - 4 \alpha_{\mu \mu} )$ factor in eq.~\eqref{P-mumu-alpha-1st}, an overall factor. As it can be absorbed into the flux normalization uncertainty, it is unlikely that this factor significantly affects the result of $s^2_{24}$ bound. Moreover we have observed just above that the near-far two detector setting allows them to discriminate between the oscillatory effect and a constant terms.\footnote{
The MINOS analysis does contain the atmospheric-scale oscillations, and it appears that this term plays an important role in the analysis. If we engage an extended MINOS analysis with the factor $( 1 - 4 \alpha_{\mu \mu} )$, one may wonder whether this factor is universal to the $\nu$SM atmospheric-scale oscillations, not only in the $\Delta m^2_{41}$-driven sterile oscillations. Fortunately, the same factor $( 1 - 4 \alpha_{\mu \mu} )$ exists also in the $\nu$SM part as an overall normalization factor, and hence our above argument is valid. }
Furthermore, $\alpha_{\mu \mu}$ is small, bounded by a few times $10^{-2}$, as we learn in section~\ref{sec:alpha-ee-mumu-mue}. 

Bringing back the above ignored $\widetilde{\alpha}_{\mu e}$ term does not alter the conclusion. The term is proportional to $s_{24} | \widetilde{\alpha}_{\mu e} | \sin 2\theta_{14}$. In section~\ref{sec:diagonal-alpha} we will learn that $| \widetilde{\alpha}_{\mu e} | \leq$ a few times $10^{-2}$, and hence $s_{24} | \widetilde{\alpha}_{\mu e} |$ is of the order of $\lsim 10^{-3}$. This shows that our above treatment is consistent with the $\alpha$ parameter effect only in the overall factor, which is to be renormalized to an over-all uncertainty, leaving the MINOS/MINOS+ bound on $s^2_{24}$ intact. 

\subsection{Possibility of MINOS/MINOS+ bound on $\alpha_{\mu \mu}$} 
\label{sec:A-mumu-MINOS} 

To obtain the $\alpha_{\mu \mu}$ bound from the MINOS/MINOS+ data, we have to derive the constraint on $s^2_{25}$ in the simplest possible setting $N_{s} = 1$ of the Okubo's method. Then, we need to treat, in addition to the two-frequency system with the atmospheric $\Delta m^2 \sim10^{-2}$ eV$^2$ and the first-sterile $\Delta m^2 \sim (1-10)$ eV$^2$, averaged-out $\Delta m^2 \gsim100$ eV$^2$ oscillations, 
They say that this highest frequency oscillation is averaged out in the both detectors, leaving a constant effect which may mix with the overall normalization uncertainties. Remember that the first sterile oscillation is averaged out in the far detector, but {\em not} in the near detector. Therefore, it is highly unlikely that accuracy of constraining $s^2_{25}$ is comparable with that of $s^2_{24}$. The task of pursuing this line further can be  carried out only by the MINOS collaboration, which we would like to gratefully encourage. 

\section{The Okubo construction} 
\label{sec:Okubo-construction} 

In section~\ref{sec:okubo-brief} we have introduced the Okubo's construction of a unitary $n \times n$ matrix, $U^{n \times n}$ by using the elements $\omega_{ij}$, see eq.~\eqref{U-nxn}. $\omega_{ij}$ denotes the $n \times n$ unit matrix apart from the replacement of the $ij$ subspace by the following $2 \times 2$ rotation matrix with the angle $\theta_{ij}$ and the phase $\phi_{ij}$: 
\begin{eqnarray} 
&&
\left[
\begin{array}{cc}
\cos \theta_{ij} & \sin \theta_{ij} e^{- i \phi_{ij} } \\
- \sin \theta_{ij} e^{ i \phi_{ij} } & \cos \theta_{ij} \\
\end{array}
\right]. 
\nonumber 
\end{eqnarray}

In section~\ref{sec:okubo-brief}, we have discussed only the $\alpha$ matrix which parametrize non-unitarity in the $(3+1)$ model. For the spirit of pedestrian approach and pedagogy, we start here from the non-unitary $\nu$SM. 

\subsection{$\alpha$ parameters in the non-unitary $\nu$SM}
\label{sec:alpha-nuSM}

If we make a decomposition 
\begin{eqnarray} 
&&
U^{n \times n} = U^{n - N} U^{N}, 
\label{decomposition}
\end{eqnarray}
for the case $n=6$, $N=3$, we obtain the non-unitary $\nu$SM. $U^{6 - 3}$ and $U^{3}$ are given by 
\begin{eqnarray} 
&&
U^{6 - 3} = 
\omega_{56} \omega_{46} \omega_{36} \omega_{26} \omega_{16} 
\cdot 
\omega_{45} \omega_{35} \omega_{25} \omega_{15} 
\cdot 
\omega_{34} \omega_{24} \omega_{14}, 
\nonumber \\
&&
U^{3} 
= 
\omega_{23} \omega_{13} \cdot \omega_{12}. 
\label{U63-U3}
\end{eqnarray}
It is informative to give the explicit matrix forms of the three parts of $U^{6 - 3}$ by using the notation $\hat{s}_{ij} \equiv s_{ij} e^{ - i \phi_{ij} }$ and $\hat{s}_{ij}^* \equiv s_{ij} e^{ i \phi_{ij} }$: 
\begin{eqnarray} 
\omega_{56} \omega_{46} \omega_{36} \omega_{26} \omega_{16} 
&=& 
\left[
\begin{array}{cccccc}
c_{16} & 0 & 0 & 0 & 0 & \hat{s}_{16} 
\\
- \hat{s}_{26} \hat{s}_{16}^* & c_{26} & 0 & 0 & 0 & \hat{s}_{26} c_{16} \\
- \hat{s}_{36} c_{26} \hat{s}_{16}^* & 
- \hat{s}_{36} \hat{s}_{26}^* & 
c_{36} & 0 & 0 & \hat{s}_{36} c_{26} c_{16} \\
- \hat{s}_{46} c_{36} c_{26} \hat{s}_{16}^* & 
- \hat{s}_{46} c_{36} \hat{s}_{26}^* & 
- \hat{s}_{46} \hat{s}_{36}^* & 
c_{46} & 0 & \hat{s}_{46} c_{36} c_{26} c_{16} \\
- \hat{s}_{56} c_{46} c_{36} c_{26} \hat{s}_{16}^* & 
- \hat{s}_{56} c_{46} c_{36} \hat{s}_{26}^* & 
- \hat{s}_{56} c_{46} \hat{s}_{36}^* & 
- \hat{s}_{56} \hat{s}_{46}^* & 
c_{56} & \hat{s}_{56} c_{46} c_{36} c_{26} c_{16} \\
- c_{56} c_{46} c_{36} c_{26} \hat{s}_{16}^* & 
- c_{56} c_{46} c_{36} \hat{s}_{26}^* & 
- c_{56} c_{46} \hat{s}_{36} ^* & 
- c_{56} \hat{s}_{46} ^* & 
- \hat{s}_{56} ^* & c_{56} c_{46} c_{36} c_{26} c_{16} \\
\end{array}
\right]. 
\nonumber \\
\label{U(6-3)-56}
\end{eqnarray}
\begin{eqnarray} 
\omega_{45} \omega_{35} \omega_{25} \omega_{15} 
&=&
\left[
\begin{array}{cccccc}
c_{15} & 0 & 0 & 0 & \hat{s}_{15} & 0 \\
- \hat{s}_{25} \hat{s}_{15} ^* & 
c_{25} & 0 & 0 & \hat{s}_{25} c_{15} & 0 \\
- \hat{s}_{35} c_{25} \hat{s}_{15}^* & 
- \hat{s}_{35} \hat{s}_{25} ^* & c_{35} & 0 & 
\hat{s}_{35} c_{25} c_{15} & 0 \\
- \hat{s}_{45} c_{35} c_{25} \hat{s}_{15} ^* & 
- \hat{s}_{45} c_{35} \hat{s}_{25} ^* & 
- \hat{s}_{45} \hat{s}_{35} ^* & c_{45} & 
\hat{s}_{45} c_{35} c_{25} c_{15} & 0 \\
- c_{45} c_{35} c_{25} \hat{s}_{15} ^* & 
- c_{45} c_{35} \hat{s}_{25} ^* & 
- c_{45} \hat{s}_{35} ^* & - \hat{s}_{45} ^* & 
c_{45} c_{35} c_{25} c_{15} & 0 \\
0 & 0 & 0 & 0 & 0 & 1 \\
\end{array}
\right].
\label{U(6-3)-45}
\end{eqnarray}
\begin{eqnarray} 
\omega_{34} \omega_{24} \omega_{14} 
&=&
\left[
\begin{array}{cccccc}
c_{14} & 0 & 0 & \hat{s}_{14} & 0 & 0 \\
- \hat{s}_{24} \hat{s}_{14}^* & c_{24} & 0 & \hat{s}_{24} c_{14} & 0 & 0 \\
- \hat{s}_{34} c_{24} \hat{s}_{14}^* & - \hat{s}_{34} \hat{s}_{24}^* & c_{34}  & \hat{s}_{34} c_{24} c_{14} & 0 & 0 \\
- c_{34} c_{24} \hat{s}_{14}^* & - c_{34} \hat{s}_{24}^* & - \hat{s}_{34}^* & c_{34} c_{24} c_{14} & 0 & 0 \\
0 & 0 & 0 & 0 & 1 & 0 \\
0 & 0 & 0 & 0 & 0 & 1 \\
\end{array}
\right]. 
\label{U(6-3)-34}
\end{eqnarray}
Notice that the standard $\nu$SM mixing matrix is buried into the upper-left $3 \times 3$ sub-matrix in $U^{3}$ as 
\begin{eqnarray}
U^{3} = 
\left[
\begin{array}{cccccc}
U^{3}_{11} & U^{3}_{12} & U^{3}_{13} & 0 & 0 & 0 \\
U^{3}_{21} & U^{3}_{22} & U^{3}_{23} & 0 & 0 & 0 \\
U^{3}_{31} & U^{3}_{32} & U^{3}_{33} & 0 & 0 & 0 \\
0 & 0 & 0 & 1 & 0 & 0 \\
0 & 0 & 0 & 0 & 1 & 0 \\
0 & 0 & 0 & 0 & 0 & 1 \\
\end{array}
\right] 
\label{U3}
\end{eqnarray}
Then, it is obvious that the similar upper-left $3 \times 3$ sub-matrix $U^{6 - 3}$, see eq.~\eqref{U63-U3}, produces the $\alpha$ matrix. By carrying out multiplication  of the three parts given in eqs.~\eqref{U(6-3)-56} - \eqref{U(6-3)-34}, the similar upper-left $3 \times 3$ sub-matrix can be parametrized as 
\begin{eqnarray}
&& 
U^{6 - 3} \vert_{3 \times 3} 
=
\left[
\begin{array}{ccc}
( 1 - \alpha_{ee} ) & 0 & 0 \\
- \alpha_{\mu e} & ( 1 - \alpha_{\mu \mu} ) & 0 \\
- \alpha_{\tau e} & - \alpha_{\tau \mu} & ( 1 - \alpha_{\tau \tau} ) \\
\end{array}
\right] 
\equiv 
1 - \alpha_{\text{\tiny (3x3)}}. 
\label{alpha-nuSM}
\end{eqnarray}
Then, the $\alpha$ parameters in the non-unitary $\nu$SM are given by 
\begin{eqnarray} 
&&
( 1 - \alpha_{e e} ) 
= 
c_{16} c_{15} c_{14} 
\nonumber \\
&&
( 1 - \alpha_{\mu \mu} ) 
= 
c_{26} c_{25} c_{24} 
\nonumber \\
&&
( 1 - \alpha_{\tau \tau} ) 
= 
c_{36} c_{35} c_{34} 
\nonumber \\
&&
\alpha_{\mu e } 
= 
\hat{s}_{26} \hat{s}_{16}^* c_{15} c_{14} 
+ c_{26} \left( \hat{s}_{25} \hat{s}_{15} ^* c_{14} + c_{25} \hat{s}_{24} \hat{s}_{14}^* \right) 
\nonumber \\
&&
\alpha_{\tau e} 
= 
\hat{s}_{36} c_{26} \hat{s}_{16}^* c_{15} c_{14} 
- \hat{s}_{36} \hat{s}_{26}^* \left( \hat{s}_{25} \hat{s}_{15} ^* c_{14} + c_{25} \hat{s}_{24} \hat{s}_{14}^* \right) 
+ c_{36} \left( \hat{s}_{35} c_{25} \hat{s}_{15}^* c_{14} - \hat{s}_{35} \hat{s}_{25} ^* \hat{s}_{24} \hat{s}_{14}^* + c_{35} \hat{s}_{34} c_{24} \hat{s}_{14}^* \right) 
\nonumber \\
&&
\alpha_{\tau \mu} 
= 
\hat{s}_{36} \hat{s}_{26}^* c_{25} c_{24} + c_{36} \left( \hat{s}_{35} \hat{s}_{25} ^* c_{24} + c_{35} \hat{s}_{34} \hat{s}_{24}^* \right) 
\label{alpha-elements-nuSM}
\end{eqnarray}

\subsection{$\alpha$ parameters in the non-unitary $(3+1)$ model}
\label{sec:alpha-3+1}

To obtain the non-unitary $(3+1)$ model from the same $n=6$ model, we make a different decomposition $U^{n \times n} = U^{n - N} U^{N}$ in eq.~\eqref{decomposition} but with $n=6$, $N=4$. 
This is given in eq.~\eqref{U64-U4} in section~\ref{sec:okubo-brief}. 
We note that $U^{4}$ has the two blob, $4 \times 4$ $U$ sub matrix and $2 \times 2$ unit matrix. See the similar $U^{3}$ matrix in eq.~\eqref{U3} in the non-unitary $\nu$SM. Therefore, if we focus on the upper-left $4 \times 4$ submatrix in $U^{6 - 4}$, this is nothing but the form given in $N = ( 1 - \alpha) U$ in eq.~\eqref{alpha-def}. 
\begin{eqnarray} 
&& 
U^{6 - 4} \vert_{4 \times 4} 
= 
\left[
\begin{array}{cccc}
( 1 - \alpha_{e e} ) & 0 & 0 & 0 \\
- \alpha_{\mu e } & ( 1 - \alpha_{\mu \mu} ) & 0 & 0 \\
- \alpha_{\tau e} & - \alpha_{\tau \mu} & ( 1 - \alpha_{\tau \tau} ) & 0 \\ 
- \alpha_{S e} & - \alpha_{S \mu} & - \alpha_{S \tau} & ( 1 - \alpha_{SS} ) \\
\end{array}
\right] 
\equiv 
1 - \alpha_{\text{\tiny (4x4)}}. 
\label{alpha-3+1-UV}
\end{eqnarray}
Then, 
the $\alpha$ matrix elements have explicit expressions by using $c_{ij} \equiv \cos \theta_{ij}$, 
$\hat{s}_{ij} \equiv \sin \theta_{ij} e^{ - i \phi_{ij} }$, and 
$\hat{s}_{ij}^* \equiv \sin \theta_{ij} e^{ i \phi_{ij} }$ as 
\begin{eqnarray} 
&&
( 1 - \alpha_{e e} ) 
= 
c_{16} c_{15}, 
\nonumber \\
&&
( 1 - \alpha_{\mu \mu} ) 
= 
c_{26} c_{25}, 
\nonumber \\
&&
( 1 - \alpha_{\tau \tau} ) 
= 
c_{36} c_{35}, 
\nonumber \\
&&
( 1 - \alpha_{S S} ) 
= 
c_{46} c_{45}, 
\nonumber \\
&&
\alpha_{\mu e } 
= 
( \hat{s}_{26} \hat{s}_{16}^* c_{15} + c_{26} \hat{s}_{25} \hat{s}_{15} ^* ), 
\nonumber \\
&&
\alpha_{\tau e} 
= 
( \hat{s}_{36} c_{26} \hat{s}_{16}^* c_{15} 
- \hat{s}_{36} \hat{s}_{26}^* \hat{s}_{25} \hat{s}_{15} ^* 
+ c_{36} \hat{s}_{35} c_{25} \hat{s}_{15}^* ), 
\nonumber \\
&&
\alpha_{\tau \mu} 
= 
( \hat{s}_{36} \hat{s}_{26}^* c_{25} 
+ c_{36} \hat{s}_{35} \hat{s}_{25} ^* ), 
\nonumber \\
&&
\alpha_{S e} 
= 
( \hat{s}_{46} c_{36} c_{26} \hat{s}_{16}^* c_{15} 
- \hat{s}_{46} c_{36} \hat{s}_{26}^* \hat{s}_{25} \hat{s}_{15} ^* 
- \hat{s}_{46} \hat{s}_{36}^* \hat{s}_{35} c_{25} \hat{s}_{15}^* 
+ c_{46} \hat{s}_{45} c_{35} c_{25} \hat{s}_{15} ^* ), 
\nonumber \\
&&
\alpha_{S \mu} 
= 
( \hat{s}_{46} c_{36} \hat{s}_{26}^* c_{25} 
- \hat{s}_{46} \hat{s}_{36}^* \hat{s}_{35} \hat{s}_{25} ^* 
+ c_{46} \hat{s}_{45} c_{35} \hat{s}_{25} ^* ), 
\nonumber \\
&&
\alpha_{S \tau} 
= 
( \hat{s}_{46} \hat{s}_{36}^* c_{35} 
+ c_{46} \hat{s}_{45} \hat{s}_{35} ^* ). 
\label{alpha-elements-3+1}
\end{eqnarray}

\end{document}